\documentclass[apj]{emulateapj}

\usepackage{natbib}
\usepackage{epsfig}
\usepackage{url}
\newcommand{\src}{AR Scorpii}
\usepackage{bm}
\usepackage{color}
\begin{document}

\title{An X-ray study for white dwarf binary AR Scorpii}

\author{J. Takata\altaffilmark{1}, X.F. Wang\altaffilmark{1}, H.H. Wang\altaffilmark{1}, L.C.-C. Lin\altaffilmark{2}, C.-P. Hu\altaffilmark{3}, K.L. Li\altaffilmark{4}, and A.K.H. Kong\altaffilmark{5}}
\altaffiltext{1}{Department of Astronomy, School of Physics, Huazhong University of Science and Technology, Wuhan 430074, China}
\altaffiltext{2}{Department of Physics, UNIST, Ulsan 44919, Korea}
\altaffiltext{3}{Department of Physics, National Changhua University of Education, Changhua 50007, Taiwan}
\altaffiltext{4}{Department of Physics, National Cheng Kung University, 70101 Tainan, Taiwan}
\altaffiltext{5}{Institute of Astronomy, National Tsing Hua University, Hsinchu 30013, Taiwan}
\email{takata@hust.edu.cn}

\begin{abstract}
 We report a study of the X-ray emission from the white dwarf/M-type star binary system \src\ using archival data taken in 2016-2020. It has been known that the X-ray emission is dominated by the optically thin thermal plasma emission, and its  flux level varies significantly over the orbital phase.  The X-ray emission also contains a component  that modulates with the beat frequency between the white dwarf's spin frequency and orbital frequency. In this new analysis, the 2020 data taken by NICER shows that the X-ray emission is modulating with the spin frequency as well as the beat frequency, indicating that part of the X-ray emission  is coming from the white dwarf's magnetosphere. It is found that the signal of the spin frequency appears only at a specific orbital phase, while the beat signal  appears over the orbital phase. We interpret the X-ray emission modulating with the spin frequency and the beat frequency  as a result of the synchrotron emission from electrons with a smaller and larger 
pitch angle, respectively. In a long-term evolution, the beat pulse profile averaged over the orbital phase  changed from a single-peak structure in 2016/2018 to a double-peak structure in 2020.  The observed X-ray flux levels  measured in 2016/2017 are higher  than those measured  in 2018/2020. The plasma temperature and  amplitude of the orbital waveform might vary with time too. These results indicate that the X-ray emission from \src\  evolves on a timescale of years. This long-term evolution would be explained by a super-orbital modulation related to, for example, a precession of the white dwarf, or a fluctuation of the system related to  activity of the companion star. 

\end{abstract}
\section{Introduction}
\src\ is a binary system composed of a white dwarf (hereafter WD) and a M-type main-sequence star, and its distance  from the Earth is estimated to be $d\sim 116$pc \citep{marsh16,bailer18,peterson19}.  The emission from radio to X-ray bands is modulating with the orbital period of $P_{o}\sim 3.56$~hours \citep{marsh16, stanway18, takata18}. \cite{marsh16} report that in addition to the orbital modulation, the radio/optical emission from \src\ is modulating in a beat frequency, $\nu_{b}=\nu_{s}-\nu_{o}\sim 8.46$~mHz, where $\nu_{s}\sim 8.54$~mHz is the spin frequency of the WD and $\nu_{o}\sim 0.078$~mHz is the orbital frequency. AR Scorpii is the first WD that shows a modulation of the radio emission  related to the WD's spin. The pulsed emission up to soft X-ray bands has been confirmed \citep{takata18}, and no   confirmation of the GeV emission has been reported \citep{kaplan19, singh20}. A spin down of the WD has been  measured by the optical observations \citep{marsh16,potter18a, stiller18, gaibor20}. With a longer base line of the optical observations, \cite{gaibor20} measure a spin down rate as  $\dot{\nu}_s\sim (-4.82\pm 0.18)\times 10^{-17}{\rm~Hz~s^{-1}}$,  indicating the spin down power of  $L_{sd}\sim 2\times 10^{33}{\rm~erg~s^{-1}}$. 

\cite{marsh16} report that the optical emission is minimum around the inferior conjunction of the companion star's orbit,  where the companion star is between the WD and Earth, while it achieves  the maximum level slightly earlier than the superior conjunction, where the companion star is behind the WD.  \cite{littlefield17} find that the orbital modulation has been fairly stable with 78 days of observations, but suggest with a longer base line optical data that the orbital maximum shifts toward earlier phases relative to the superior conjunction with time. \cite{katz17} interprets this  observed orbital  waveform and the long-term evolution with a precession model of the WD's spin axis.  \cite{peterson19}, on the other hand, report that the orbital waveform of \src\ has remained constant across about one century, and constrain the first time derivative of the orbital frequency as $\dot{\nu}_{o}\le 3.8\times 10^{-20} {\rm ~Hz~s^{-1}}$. 

\cite{marcote17}  carry out a radio observation with a high angular resolution and find that the radio emission region  of \src\ is compact with a scale of $\le 4R_{\odot}$.  They also report no evidence for a radio outflow or collimated jets.  Then,  \cite{stanway18} confirm that the orbital modulation in radio bands shows a similar trend to that of the optical emission. This indicates that  inner-surface  of the companion star is probably  heated by an interaction with the WD magnetosphere or by an irradiation of synchrotron radiation from  the WD's magnetosphere \citep{garnavich19}, and the emission from radio to optical bands is mainly produced at the region near or on the inner-surface. 

The pulse profile folded with the beat frequency shows a double-peak structure in the radio, optical and UV bands. In optical/UV bands, the pulse shape can be described  by a  main peak plus a small secondary peak, for which the phase separation is $\sim 0.5$ \citep{marsh16}. The pulsed fraction reaches  $>95$\% in the UV band. \cite{marsh16} also find a signal of the spin frequency of the WD in the periodograms, suggesting that  the optical pulsed emission originates  from a region co-rotating with the WD. The optical emission from \src\ is also unique with a strong linear polarization. \cite{buckley17} show that the degree of the linear  polarization reaches up to $\sim 40\%$ at the pulse peak, and the position angle swings through $180^{\circ}$. These properties of the linear polarization indicate an origin of   synchrotron emission from the relativistic electrons spiraling along the magnetic field of the WD \citep{potter18b, takata19, duplessis19}.

In radio bands, \cite{marsh16} show the pulse profile can be  described by a double-peak structure with a pulsed fraction of $\sim 10\%$, which is smaller than those in the optical/UV bands.  \cite{stanway18} report  that the  beat signal decreases with  the radio frequency, and no beat signal is found  at 1.5GHz. The radio emission does not show an evident signal of the WD spin in the periodograms, which indicates that the pulsed radio emission originates  from the region  in the magnetosphere of the companion star rather than in the magnetosphere of the WD. \cite{stanway18} also measure  a weak linear polarization with a degree $<1\%$ but a large circular polarization, which reaches to  $\sim 30$\% at the orbital phase $\phi_{o}\sim 0.8$, where $\phi_{o}=0$ is defined at the inferior conjunction of the companion star's orbit.  They suggest that a non-relativistic cyclotron emission dominates the radio emission at lower frequencies.

The broadband spectrum from the radio to UV bands can be described  by a thermal optical emission from the companion star plus non-thermal component with a total luminosity of $L\sim 2\times 10^{32}{\rm~erg ~s^{-1}}$, which is about 10\% of the spin down power. \cite{garnavich19} measure the slope of the optical spectrum and find that  the power-law index $s$, $F_{\nu}\propto \nu^{-s}$, shows a significant variation in $0.8<s<2.0$ depending on  the orbital phase and  the optical brightness. They suggest that  the variation in the power-law index is related to the synchrotron cooling and frequency of the  replenishment of the energy injection. This non-thermal component has an energy peak at around 0.01eV and probably extends down to radio bands with a slope of $F_{\nu}\sim \nu^{1/3}$ \citep{stanway18}, which is consistent with the slope of low energy tail of the synchrotron emission \citep{geng16}. In the radio bands, the spectrum becomes shallower below $\sim 1$GHz \citep{marsh16, stanway18}. This also  indicates that an additional component (probably cyclotron emission from non-relativistic electrons) dominates at low radio frequency bands.

 In the higher energy band,  we \citep{takata18} study the X-ray emission from  \src\ with an XMM-Newton observation carried out in 2016.  The observed spectrum  shows an iron-K line that  is well described by an optically thin thermal plasma emission with several different temperatures; there is no signature of the emission from an accreting matter in the observed spectrum.  The observed X-ray flux varies significantly over the orbital phase (Figure~\ref{light-orb}), and 
  it achieves  maximum and minimum at around  the superior conjunction and inferior conjunction of the companion star, respectively. The hardness ratio of the observed X-ray emission, on the other hand, is almost constant over the orbital phase.  These emission properties are compatible to the  thermal process of the plasmas heated to $0.1-10$ keV on or near the inner surface of the companion star, and  the orbital variation of the observed flux is caused by a variation of the visible emission region over the orbital phase rather than an absorption by the accretion column. The X-ray luminosity is of the order of $L_X\sim 4\times 10^{30}~\rm{erg~s^{-1}}$, and it is $\sim 1$\% of the optical/UV luminosity of the non-thermal component.  Hence, a  small
 fraction of the released energy   is used for heating up the plasma to 
 a temperature of $\sim$keV, and majority is for creating a population of the non-thermal electrons. 
 
\cite{takata18}  find a modulation with the beat frequency  in the X-ray data below  $2$~keV.  The pulse profile averaged over the orbital phase  shows a single-peak (or probably double-peak with a small secondary peak). On the other hand, the orbital resolved analysis shows that the pulse structure evolves over the orbital phase, and the pulse profile at around the inferior conjunction of the companion star can be described by  a double-peak structure with a prominent secondary peak. The averaged pulsed fraction is about $14\%$, suggesting  that the unpulsed component is the main component in the X-ray bands. By comparing the pulse profile measured by the EPIC camera with the profile by the  OM camera onboard XMM-Newton, we find that the X-ray peak and optical/UV peak are in phase.  This suggests that  the non-thermal component in optical/UV bands extends to the soft X-ray bands. The phase-resolved spectrum of the pulsed component in the 0.15-2~keV band shows evidence of the power-law component with a photon  index $\Gamma=2.3\pm 0.5$ and an unabsorbed flux $F_{\rm{0.15-2}}\sim 4\times 10^{-13}{\rm erg~cm^{-2}~s^{-1}}$.

 After the discovery of the radio/optical pulsed emission from  \src\ in 2016, more  X-ray data have been collected. It enables us to study a long-term evolution of the X-ray emission from \src. In this paper, we  study the X-ray emission properties of \src\ using archival data taken by XMM-Newton, Chandra, NuSTAR and NICER during 2016-2020 (Table~1).  In section~\ref{reduction}, we summarize  the  data reduction for each observation. We present the results of the timing analysis in section~\ref{timing} and of spectral analysis in section~\ref{spana}. In section~\ref{discuss}, we discuss  the X-ray emission process in the WD magnetosphere based on the results of  the study.

\section{Data reduction}
\label{reduction}
\begin{deluxetable}{lcccc}
\tablecaption{Journal of the observation data used in this study.}
\tablewidth{0pt}
\tablehead{
             & Obs. ID     & Date         & Exposure & \\
& & & (ks) &  }
\startdata
XMM &  0783940101           &      2016 Sep. 10        &     38.7    & X16      \\       
   -Newton       &       0795720101      &      2018 Feb. 19        &     30.0   & X18  \\                            
Chandra     & 19711 & 2017 June 23 & 26.1   & Ch17      \\                                                               
NuSTAR    & 30301027002 & 2018 Feb. 18 & 134.7    & Nu18     \\                                                          
NICER     &      3589010101       &       2020 June 14      &      5.6      &   \\                                        
           &  3589010102           &    2020 June 14          &    19.3      &     \\                                     
           &  3589010103           &    2020 June 16          &    15.4      &  Ni20   \\                                 
           &  3589010104           &    2020 June 16          &    9.5        &    \\                                     
           &  3589010105           &    2020 June 18          &    6.9        &    \\
\enddata
\end{deluxetable}

\subsection{$XMM$-Newton}
\label{xmm}
We analyze  the archival XMM-Newton data  (MOS1/2 and PN) taken at 2016 September 10 (Obs. ID:0783940101, PI: Steeghs) and at 2018 February 19. (Ob. ID: 0795720101, PI: Pavlov) with  total exposure times of 39~ks and 30~ks, respectively (Table~1). The observations were operated under the small window mode for the MOS1/MOS2 and the large window mode for the PN  in 2016 (hereafter X16), while with full frame mode for all EPIC cameras in 2018 (hereafter X18).  Event list for each detector and observation is produced in the standard way using  XMM-Newton Science Analysis Software (\verb|XMMSAS|, version 18.0.0). A point source is significantly detected by the  \verb|XMMSAS|  task  \verb|detect_chain|. We generate the source spectra  within a radius of $20''$ circle centered at  source position (R.A., Dec.)=($16^{\rm{h}}21^{\rm{m}}47^{\rm{s}}$.28,$-22^{\circ}53'10''.4$) and the background spectra from a source free region. When we generate the spectra, we remove epoch with background flaring by limiting 
the count rate above 10~keV to $<0.35~{\rm count~s^{-1}}$ for the MOS1/MOs2  data and  $<0.4~{\rm count~s^{-1}}$ for the PN data.  We create response files with the \verb|XMMSAS| tasks \verb|rmfgen| and \verb|arfgen| with updated calibration files. We group the spectral bins  to achieve a signal-to-noise ratio of  $S/N\ge 3$ in each bin using the \verb|XMMSAS| task \verb|specgrp|. We use the  \verb|XMMSAS| task \verb|barycen| to obtain barycenter corrected arrival times (ephemeris DE405) and create the background subtracted light curve with the task \verb|epiclccorr| to examine the orbital variation. We generate the orbital light curves in the 0.3-10keV band, and compare it with the results of the  Chandra and NICER data (Figure~\ref{light-orb}).

The  X18 data recorded  a large flare-like background event covering a significant part of the observation. After removing such a background event, the exposure of the data used in the spectral analysis is about $10$~ks for the PN data and $20$~ks for the MOS1/MOS2 data, respectively.  Figure~\ref{exposure} shows the orbital phase  distribution of the exposure time for the data used in the spectral analysis; the maximum exposure time is normalized to unity. We can see that a large exposure loss because of  the background events causes a non-uniform orbital coverage of the X18 data. For the X16 data, the exposure loss due to  the flare-like background events is less serious compared to the case of the X18 data.

\subsection{Chandra}
A Chandra observation for \src\ (Obs. ID: 19711, PI:Pavlov) was operated under ACIS Timed Exposure  (1/8 sub-array) mode and an exposure time of 24.8~ks (hereafter Ch17). We re-process the data with standard tasks in the \verb|CIAO| Version 4.11. We use the \verb|CIAO| task  \verb|specextract| to generate the  source  spectrum  and associated response files from a circular  region centered at aforementioned position with a radius $3.6''$, and create a   background spectrum from a vicinity source free region. It is known that pile-up of a Chandra data is significant when the count rate is $\geq  0.1~\rm{count~s^{-1}}$ \citep{davis01}. In section~\ref{pileup}, we examine the pile-up fraction as $\sim 5$\% for this observation. For a timing analysis, we apply the  \verb|CIAO| task \verb|axbary| to obtain  barycen-corrected arrival time of each event, and  use  \verb|dmextract| task to create  a background subtracted light curve in the  0.3-10keV band to examine the orbital variation. We check the background light curve and find  no indication of flaring.  For searching a beat frequency in the data, we use the \verb|Xselect| to extract events from the source region and obtain a photon count of $\sim$4200.

\subsection{NuSTAR}
To examine the hard X-ray emission from \src, we analyze  the archival NuSTAR data taken at 2018 February  (Obs. ID: 30301027002, PI: Pavlov) with an exposure $\sim$134~ks (hereafter Nu18). Event list and source/background spectra are produced in the standard way using  \verb|nupipline| and \verb|nuproducts| tasks of the  \verb|Heasoft|. Barycenter  correction is applied in the \verb|nuproducts| task.  A point source is clearly seen in the generated image. We create the source  spectrum with the data extracted from a circular region with a radius of $50''$ centered at aforementioned source position,  and we  obtain 
 a  background spectrum from  a vicinity source free region. We use  \verb|grppha| task of the \verb|FTOOLS| to group the spectral bins such that new grouping contains a minimum of 30 counts in each bin. We create a background subtracted light curve in the 3-78~keV band.  We  extract $\sim 10,000$ events from the source region, and  search for  the beat frequency in the data.

\subsection{NICER}
\label{nicer}
NICER observed \src\ at 2020 June 14-16 (Obs.~IDs: 589010101-358901015 PI: Mori) with a total exposure $\sim57$~ks (hereafter Ni20). We merge the \verb|mpu7_ufa| files for the five observations  using the task \verb|nimpumerge| of the \verb|Heasoft|,  and screen the data using the tasks \verb|nicerclean| and \verb|nimaketime|.  We filter the raw data with the standard criteria for analysis of the beat/spin modulation,  and we apply an additional filtering based on  the magnetic cut-off rigidity to generate  spectrum and orbital light curve. We accept the time with $>1.5$GeV/c in  current analysis and find that a more tight filtering does not affect much to the result of the analyses. We use the latest calibration files (\verb|xti20200722|) and recalibrate the energy scale using the file \verb|nixtiflightpi20170601v004.fits|. We apply a barycentric time  correction to the created event list  using  \verb|barycor| task of the \verb|Heasoft|.  After standard screening, we also remove  flare-like background events using  the \verb|Xselect|.

 We generate the spectra of the source region and background modeling with \verb|nibackgeng3C50| task implemented in the \verb|HEASoft|; we create a filter file with an option   \verb|coltypes="base,3c50"| in the task \verb|niprefilter2|\footnote{https://heasarc.gsfc.nasa.gov/lheasoft/ftools/headas/niprefilter2.html}. We  use the \verb|grppha| task to group the generated spectrum such that  new grouping contains a minimum of 40 counts in each bin. We note that the best-fit parameters for the spectrum  obtained with the original event file  is consistent with those obtained with the \verb|nibackgeng3C50| product within the errors. In this study, therefore, we will present the fitting result with the \verb|nibackgeng3C50| product.

Figure~\ref{exposure} shows that the 2020 NICER observation  does not cover uniformly the orbital phase. 
We generate an exposure corrected light curve in the 0.3-10~keV band using  the \verb|Xselect|.  
 Using the source and background spectra obtained by the \verb|nibackgeng3C50| task, we estimate that the background emission contributes to $\sim 55$\% of total emission in the 0.3-10~keV band. 
By removing  55\% of the total counts from the generated light curve,  we create a background subtracted light curve, which is compared  with other observations  (Figure~\ref{light-orb}).

\section{Timing analysis}
\label{timing}
\subsection{Properties of the orbital modulation}
\begin{figure}
 \epsscale{1}
\centerline{
\includegraphics[scale=0.5]{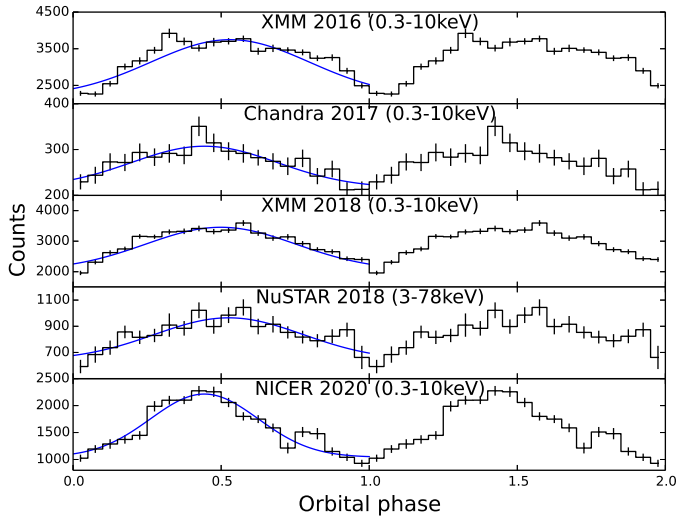}}
\caption{Orbital modulation measured by each data set.  Background contribution is subtracted and exposure loss for
 each bin is corrected (Figure~\ref{exposure}). For Ni20 data, we assume the background level as 55\% of the total
 emission in the  0.3-10~keV band (section~\ref{nicer}). The blue solid curve in each panel represents the best-fit function 
with one Gaussian component plus a constant component.}
\label{light-orb}
\end{figure}

\begin{figure}
\epsscale{1}
\centerline{
\includegraphics[scale=0.5]{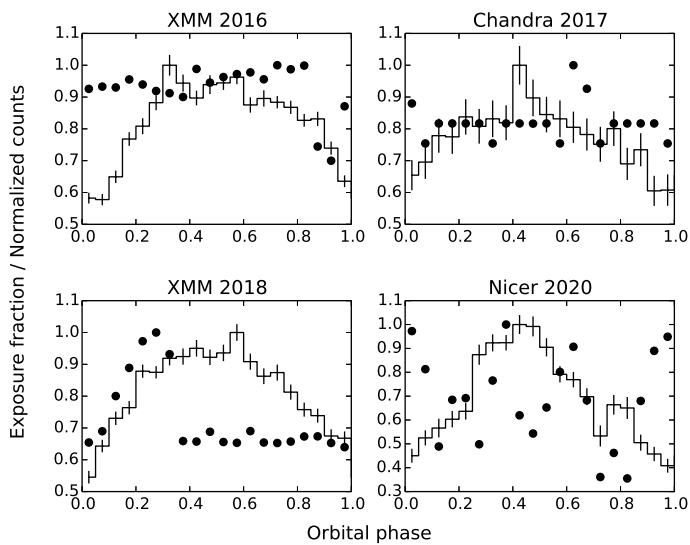}}
\caption{Distribution of the exposure fraction (filled circles) over the orbital phase. Histograms shows the orbital variation after correction of the exposure loss. For X18 data, the large exposure loss in the orbital phase $\phi_o=0.4-1$ is caused by large flare-like background events. For Ni20 data, a non-continuous monitoring of the source produces a large fluctuation of the exposure fraction over the orbital phase.}
\label{exposure}
\end{figure}

We create folded light curve with the orbital ephemeris reported by \cite{marsh16}: $\nu_{o}=0.077921380$~mHz and $T_0(\rm{MJD})=57264.09615$ when the companion star is located at the inferior conjunction.  Figure~\ref{light-orb} shows the orbital light curves in the 0.3-10~keV band for the X16/18, Ch17 and Ni20 data  and in the 3-78~keV band for the Nu18 data; the background level is subtracted for each light curve.  We can see that all light curves show a significant orbital modulation, and the observed flux levels achieve the maximum and  minimum at around the superior conjunction ($\phi_o=0.5$) and inferior conjunction ($\phi_o=0$), respectively. As discussed in Takata et al. (2018), this pattern of the orbital modulation indicates that X-ray emission from \src\  mainly originates  from the region located near or on the inner surface of the companion star.

To examine a   long-term evolution of the orbital waveform, we fit the obtained light curve with one Gaussian component plus a constant component (Figure~\ref{light-orb}). We define  amplitude of the orbital variation as $(f_{max}-f_{min})/(f_{max}+f_{min})$, where $f_{max}$ and $f_{min}$ are the maximum and minimum counts obtained from the best-fit function, respectively \citep{inam04,rea07}. We obtain  amplitudes of  $\sim 0.22\pm0.01$ for the X16 data, $\sim 0.16\pm 0.03$ for the C1h7 data, $\sim 0.21\pm 0.01$ for the X18 data and $\sim 0.18\pm 0.02$ for the Nu18 data. For the  Ni20 data, we obtain an amplitude of $\sim 0.32\pm0.01$ by assuming  $55$\%  of the background level, and  we find that the amplitude is larger than those of other data. This would provide  evidence of the long-term evolution of the orbital waveform in the X-ray 
bands, although additional observations with an  imaging capability is 
necessary to confirm it.   By comparing  the amplitude  of X18 data with that of Nu18 data, we find less dependency  of the orbital waveform on the photon energy. This constant amplitude among different energy bands also indicates that the orbital modulation is caused by a  variation of the 
visible emission region over the orbital phase. 

\subsection{Searching for pulsation}
\begin{figure}
 \epsscale{1.}
\centerline{
\includegraphics[scale=0.5]{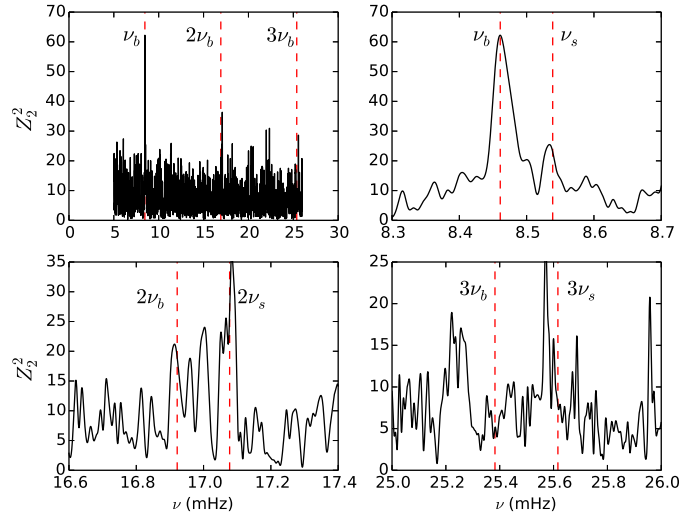}}
\caption{ $Z_2^2$ periodogram for the XMM-Newton 2018 data. $\nu_b$ and $\nu_s=\nu_b+0.0792{\rm mHz}$ represents the beat \
frequency and spin frequency of the WD, respectively.  }
\label{z18}
\end{figure}

\begin{figure}
 \epsscale{1.}
\centerline{
\includegraphics[scale=0.5]{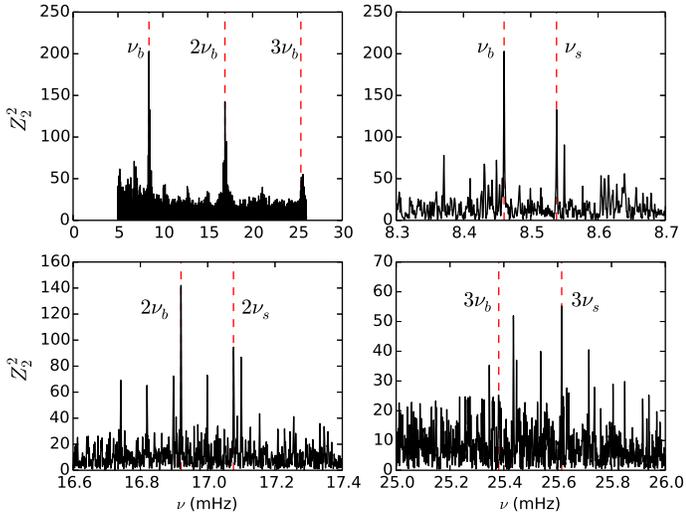}}
\caption{ $Z_2^2$ periodogram for the NICER 2020 data. $\nu_b$ and $\nu_s$ represents the 
beat frequency and spin frequency of the WD, respectively.}
\label{z}
\end{figure}

\begin{figure}
 \epsscale{1.}
\centerline{
\includegraphics[scale=0.5]{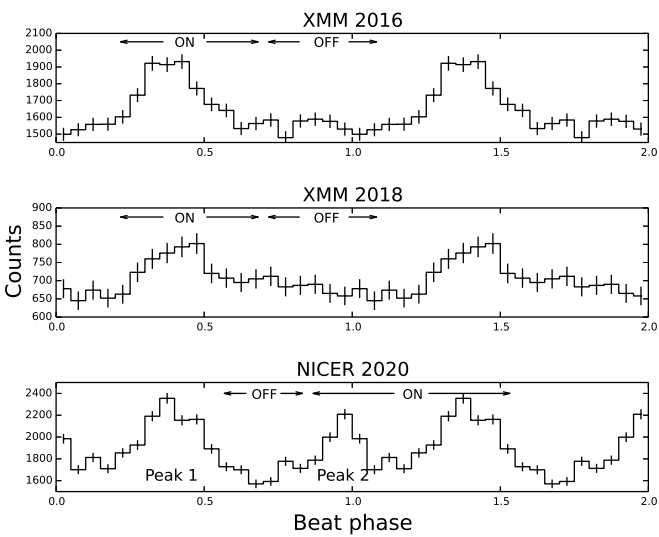}}
\caption{ Pulse profiles in the 0.3-2~keV band folded with the beat frequency for the three data sets. Each event 
is folded with the timing solution  described  in  section~\ref{longpulse}, and the background contribution is subtracted for each light curve. The 55\% of the background level in the total emission  is assumed for Ni20 data.  ``ON'' and ``OFF'' in each panel indicate 
the on-pulse phase and off-pulse phase to generate  
the spectrum of the pulsed component, respectively (section~\ref{pulspe}). }
\label{light}
\end{figure}

We search the beat frequency $\nu_b\sim 8.6$mHz \citep{marsh16} using the $Z_2^2$-test \citep{buccheri83}. For the X16 and X18 data sets, the extracted photon events from the source region yield  $\sim 45,000$ and $\sim 20,000$ counts, respectively. We use  $\sim 4,200$ counts for Ch17, $\sim 10,000$ counts for Nu18 FPMA/FPMB and $\sim 135,000$ photons for Ni20  to search the beat frequency. In addition to the X16 data reported by \cite{takata18}, a significant pulsed signal is found  in X18 and Ni20 data sets.  Since the pulsed fraction is $\sim 14\%$ \citep{takata18}, the counts of the pulsed emission in the Ch17 data will be  only $\sim 600$ or less, and it  is unlikely  to find a beat signal in the data. Although we combine 
 all observations, which include the Nu20 data,  to search a pulsation in the $>2$~keV band, we do not find any significant signal at the beat frequency. This non-detection of the beat frequency  with the number of 
the collected photon counts is expected,  because  only $\sim 2$\%   of the total emission from \src\  
 is pulsating  in the  $>2$~keV band (section~\ref{pulspe}).

 Figures~\ref{z18} and \ref{z} present the $Z_2^2$-periodograms for the X18 and Ni20 data, respectively [see \cite{takata18} for the $Z_2^2$ periodogram of the X16 data]. We can see a strong signal at the beat frequency $\nu_{b}\sim 8.46$mHz and the 1st harmonic in the Ni20 data.  In addition to the beat frequency, we can also confirm a significant signal at the spin frequency of the WD,   $\nu_{s}\sim 8.53$mHz and the 1st/2nd harmonics.  The signal of the spin frequency  with $Z_2^2\sim 125$ of the  Ni20 
data is much stronger than  $Z_2^2\sim 50$  of the  X16 data \citep{takata18} and $Z_2^2\sim 20$ of the  X18 data. The Ni20 data therefore provides a concrete  evidence  that the pulsed X-ray emission originates  from the WD magnetosphere and its emission region is co-rotating with the WD.

\subsection{Long-term evolution of pulse profile}
\label{longpulse}
\begin{figure*}
 \epsscale{1.}
\centerline{
\includegraphics[scale=0.5]{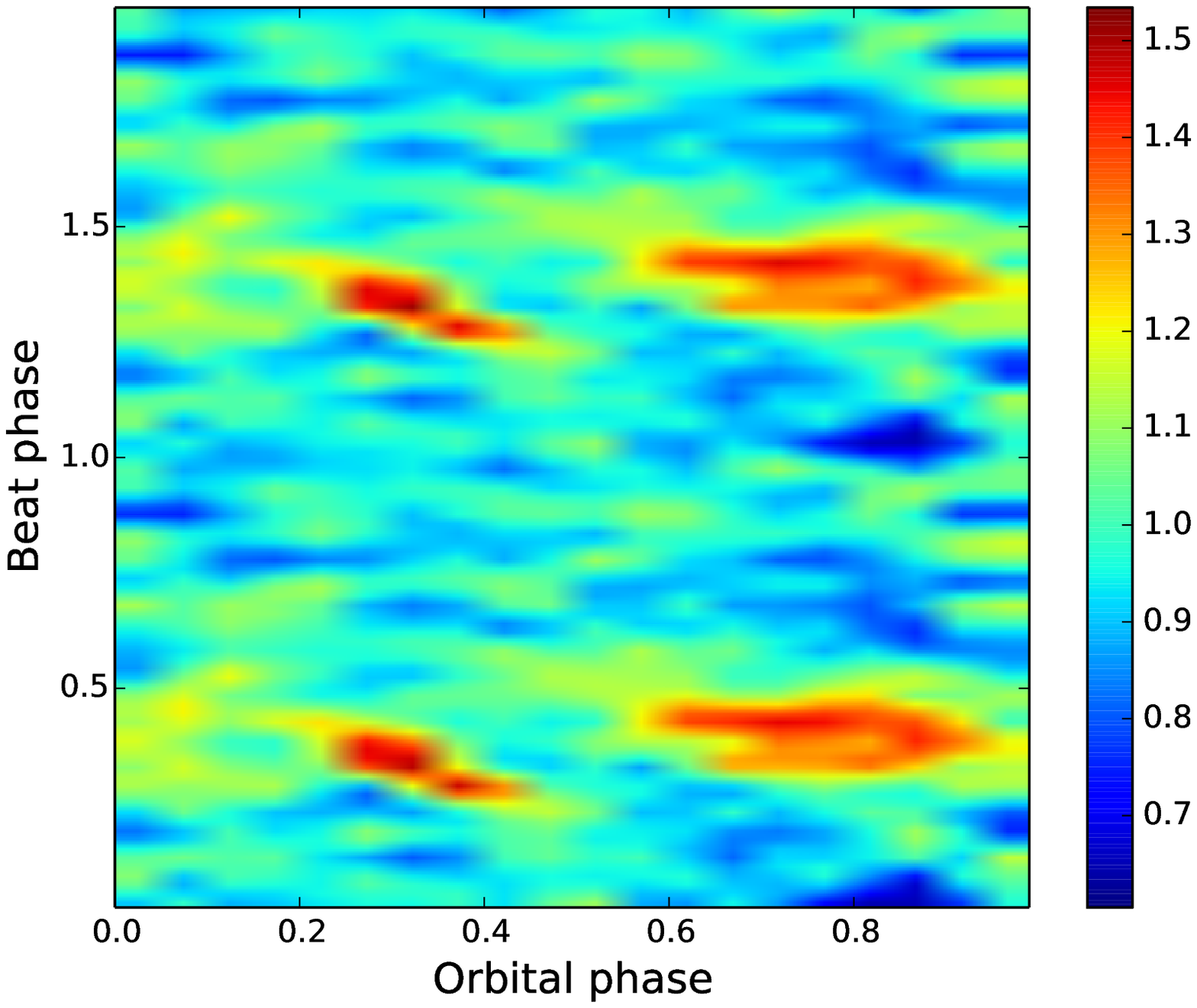}
\includegraphics[scale=0.5]{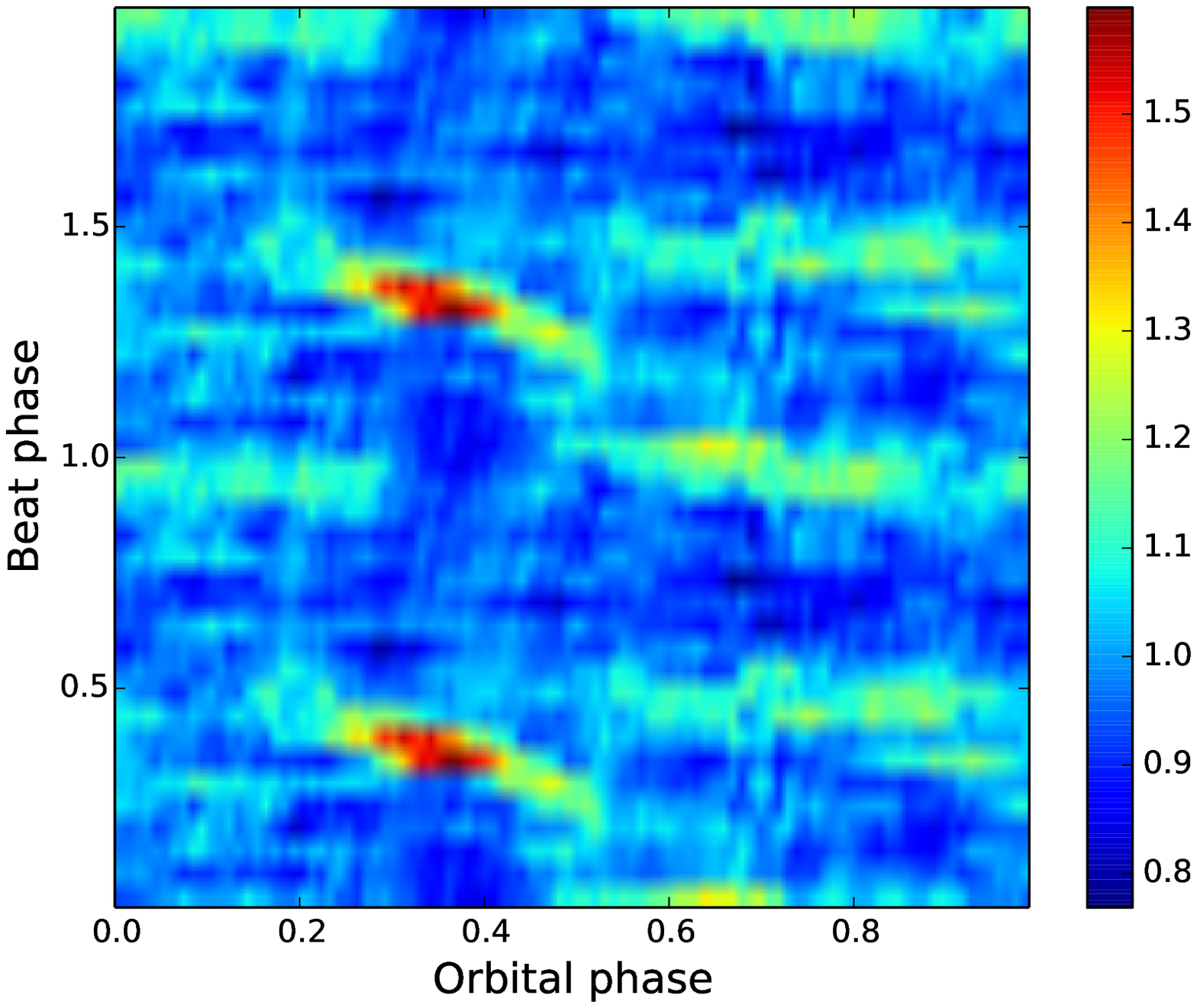}}
\caption{ Dynamic pulse profiles in the 0.3-2~keV band for X16 (left) and Ni20 (right) data. The data is folded with t\
he beat phase and the  pulse profile at each orbital phase is created by the data of successive 0.1 orbital phase.         
 Total count (area of the profile) of each pulse profile is normalized to unity.}
\label{map}
\end{figure*}
\begin{figure*}
 \epsscale{1.4}
\centerline{
\plottwo{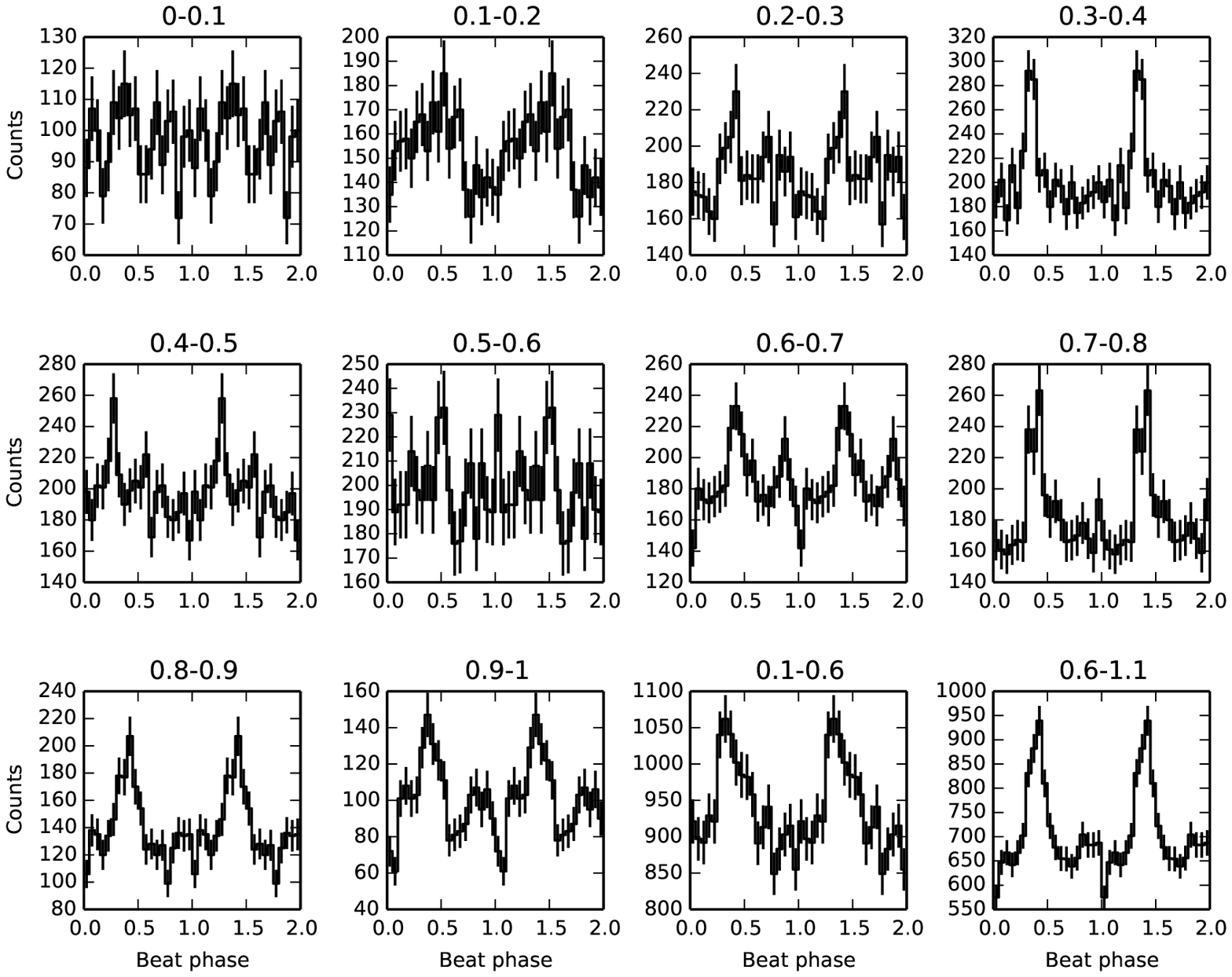}{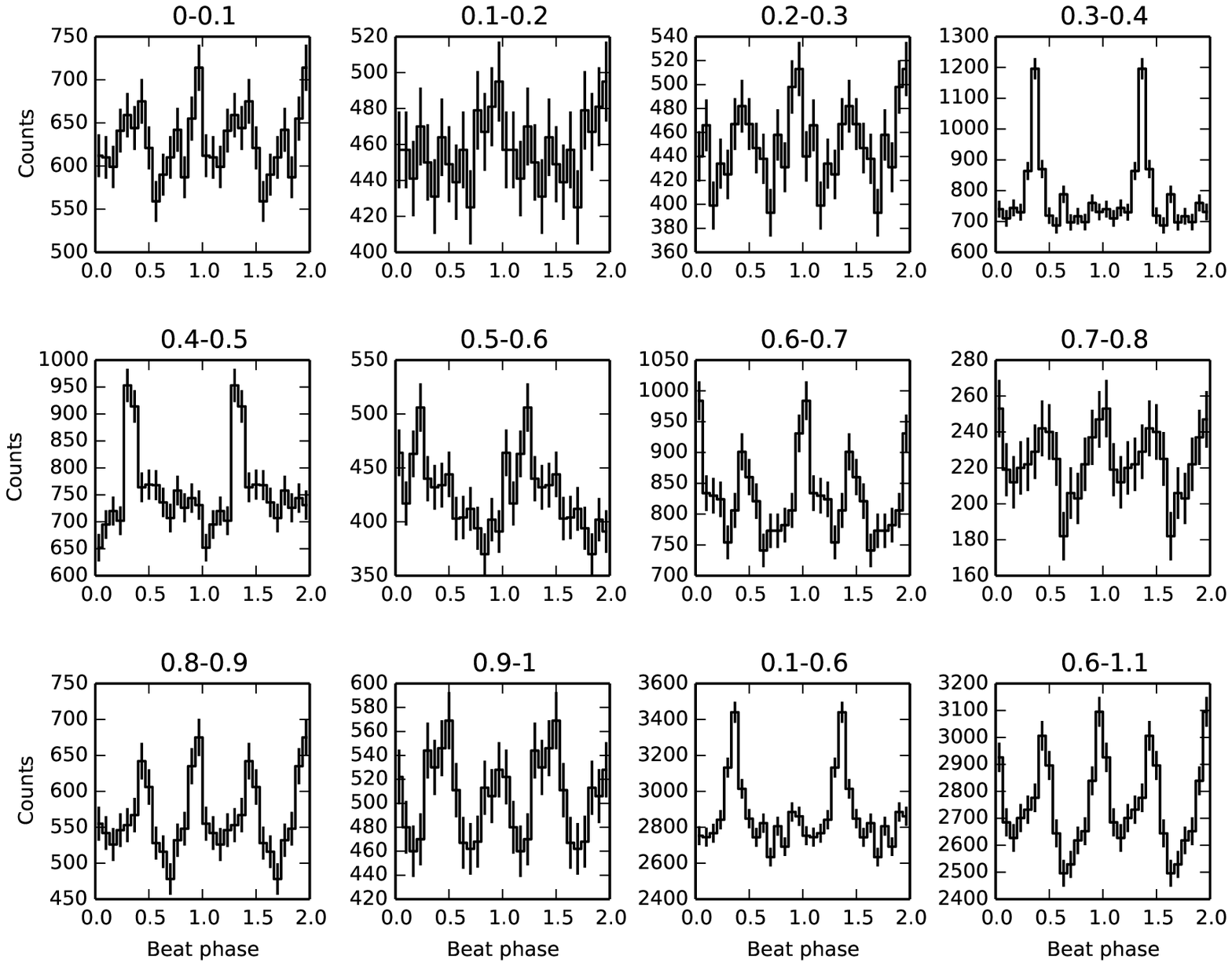}}
\caption{Orbital resolved pulse profiles in the 0.3-2~keV band for X16 (left) and Ni20 (right) data. Bottom right two \
panels show the pulse profiles in the phase segments $\phi_{o}=0.1-0.6$ and $0.6-1.1$, respectively.}
\label{ob}
\end{figure*}

\begin{figure*}
 \epsscale{1}
\centerline{
\includegraphics[scale=0.5]{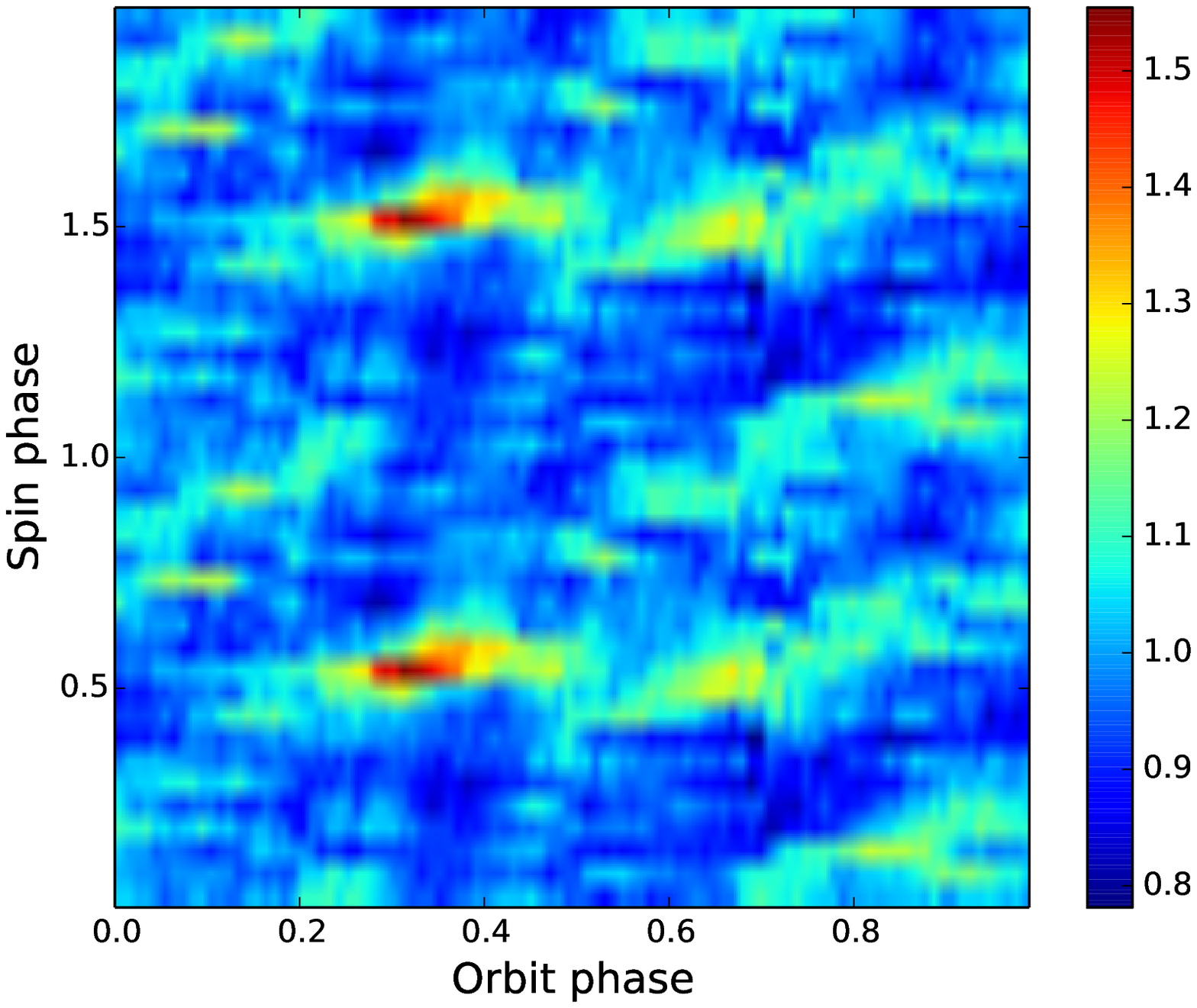}
\includegraphics[scale=0.5]{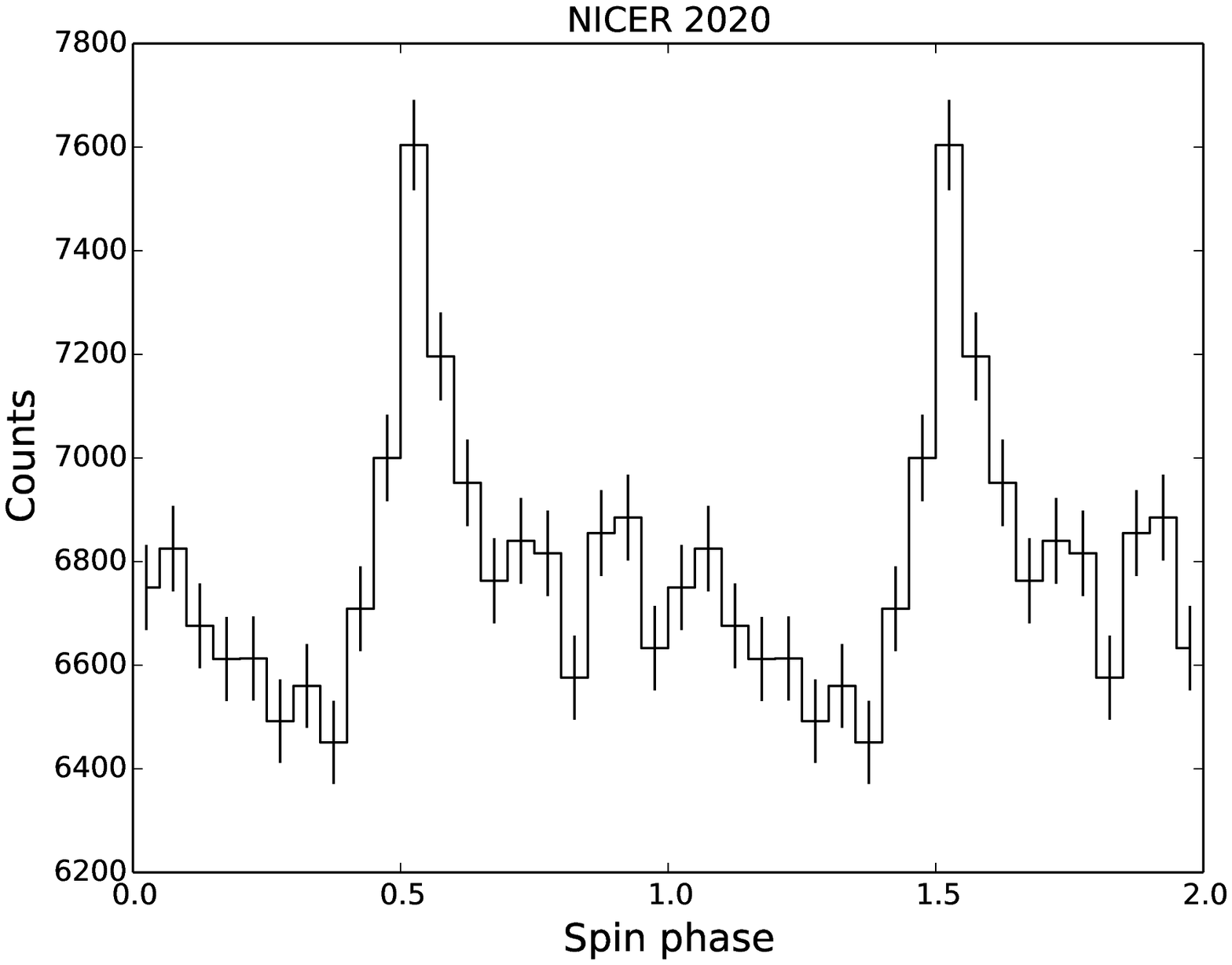}}

\caption{Left: Dynamic pulse profile folded with the spin frequency  for Ni20 data set.
Right: Pulse profiles in the 0.3-2~keV band folded with the spin frequency for Ni20 data set.}
\label{spin}
\end{figure*}

To compare between the pulse profiles measured by XMM-Newton and NICER observations, we restrict the photon energy in the  0.3-2~keV band, and create folded light curves with  the beat  ephemeris provided by \cite{gaibor20}: $\nu_{b}=8.460297018$mHz, $\dot{\nu}_{b}=-4.82\times 10^{-17}{\rm Hz~s^{-1}}$ and $T_0(\rm {MJD})=57941.16888790$.   For X16/18 data, we determine the background level from  a nearby source free region.  For Ni50 data, we generate the background subtracted light curve by assuming that the background level is 55\% of the total emission.

Figure~\ref{light} 
shows the pulse profiles for the three data sets, and we find that the pulse profile of the Ni20 data is significantly different from those of the X16/18 data. The pulse profiles of the X16/X18 data show a single-peak structure.  For the N20 data, on the other hand, the pulse profile shows a double-peak structure with a phase separation of $\delta\phi_b\sim 0.4-0.5$. The difference in the pulse profiles between the  X16/X18 and Ni20 data  may indicate  a  temporal evolution of the pulsed X-ray emission. \cite{takata18}, on the other hand, find that the pulse profile  evolves over  the orbital phase, and it can be  represented by either a single peak structure or a double-peak structure.  The pulse profile obtained with all data of each observation therefore may  depend on how the observation covers the orbital phase. In fact, we can see a large fluctuation of the  exposure time over  the orbital phase for the Ni20 data (Figure\ref{exposure}).

We examine the  orbital evolution of the pulse profiles to investigate the cause of the difference between  the pulse profiles of the X16/18 and Ni20 data.  Figure~\ref{map} is a dynamic pulse profile folded with the beat frequency over the orbital phase using the X16 data (left panel) and the  Ni20 data (right panel); the pulse profile at each orbital phase is created with  the data of successive 0.1 orbital phase, and the total count (area of the profile) is normalized to unity.  We can see that the dynamic pulse profiles in two data sets are  similar to each other,  except for feature of the secondary peak   appeared at the beat phase $\phi_{b}\sim 0$  (equivalently $\phi_{b}\sim 1$) and during the  orbital phase $\phi_{o}\sim 0.5-1$. 

 Figure~\ref{ob} shows the orbital resolved pulse profiles for the X16 (left) and Ni20 (right) data. 
In both data sets, the pulse profile at the orbital phase $\phi_{o}\sim 0.3$ can be described  by a single-peak structure appeared  at the beat phase $\phi_{b}\sim 0.5$. Then the peak position migrates to earlier beat phase during the  orbital phase $\phi_{o}\sim 0.3-0.6 $, as the dynamic pulse profiles indicate. After the orbital phase $\phi_{o}>0.6$,  we can see a double peak structure in both the X16 and N20 data. For the X16 data, the original peak at the beat phase  $\phi_{b}\sim 0.5$ always dominates the secondary peak at 
$\phi_{b}\sim 0$. For Ni20 data, on the other hand, the secondary peak has a similar or larger intensity compared to the original peak.   In the bottom row of the figure, we present the pulse profiles integrated over the orbital phase  $\phi_{o}=0.1-0.6$ and $\phi_{o}=0.6-1.1$, respectively. We find in the figure that the pulse profiles in $\phi_{o}=0.1-0.6$ for the  X16 and Ni20 both can be described  by a single-peak structure. For $\phi_{o}=0.6-1.1$, on the other hand, the pulse profile of the  Ni20 data shows  a double-peak structure 
 with a secondary peak stronger than the original  peak. For the  Ni20 data,  we can see that the  pulse profile averaged over the orbital phase  using the normalized pulse profiles  presented in  Figure~\ref{map} is also described by  the double-peak structure. These results therefore suggest that the difference between the pulse profiles of the  X16/X18 data and Ni20 data is an intrinsic feature, and the pulsed X-ray emission experiences a long-term evolution.

We define the pulsed fraction with the equation, $(f_{max}-f_{min})/(f_{max}+f_{min})$, with $f_{max}$ and $f_{min}$ being the maximum and minimum counts in the light curve. In 0.3-2.0~keV bands, we obtain $0.13\pm0.02$ for X16, $0.11\pm 0.03$ for X18, and $0.20\pm 0.02$ for Ni20 data, respectively. We find that the pulsed fraction of the Ni20 data is higher than those of the  X16/X18 data. The pulsed fraction  may show a long-term evolution, as well as the structure of the pulse profile. Although we could not find  the pulsed emission above 2~keV bands, we estimate  a pulsed fraction  less than 2\%  on the basis of  the spectral analysis (section~\ref{pulspe}). 
 
\subsection{Mixture between spin and beat modulations}

\begin{figure*}
 \epsscale{1.4}
\centerline{
\plottwo{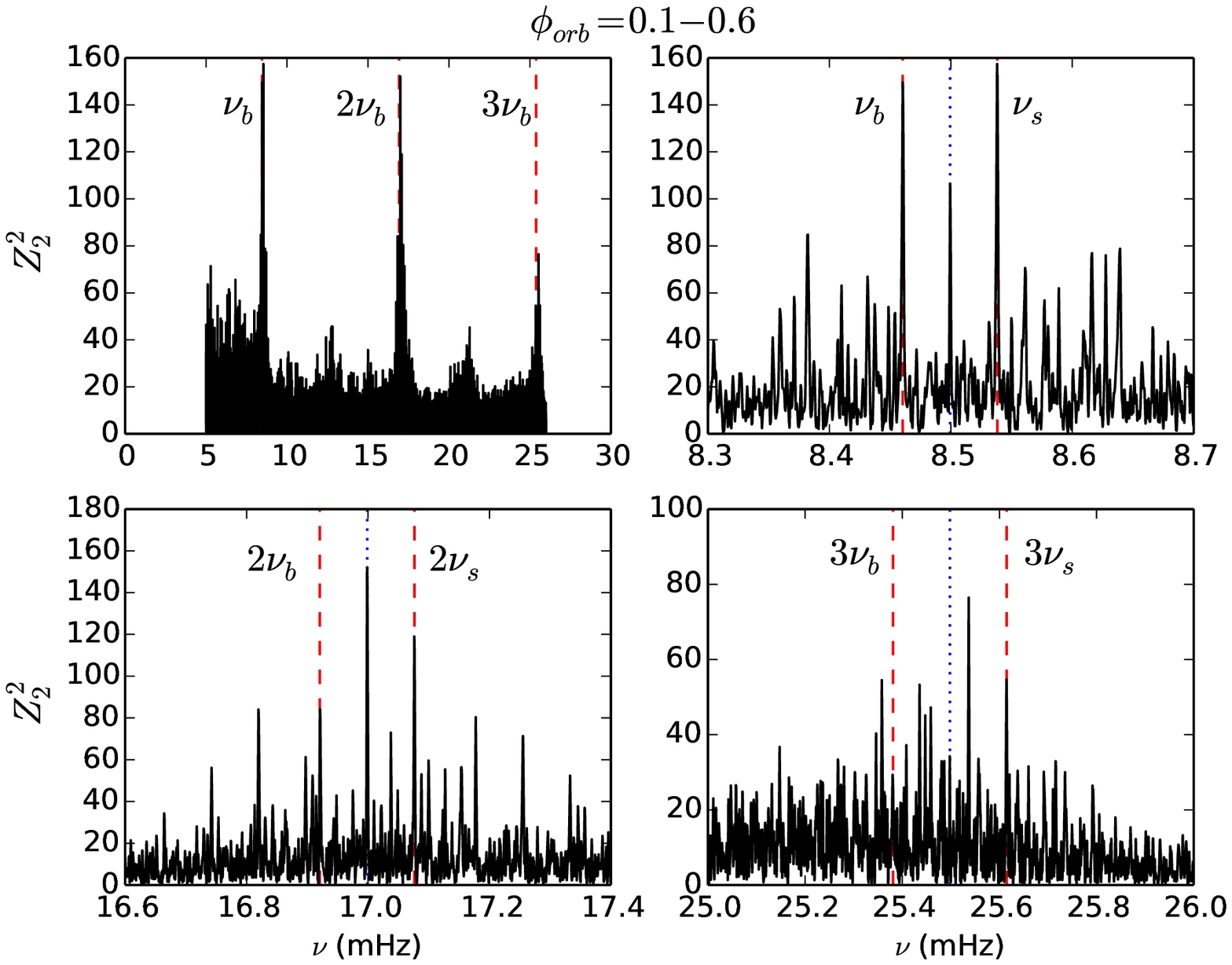}{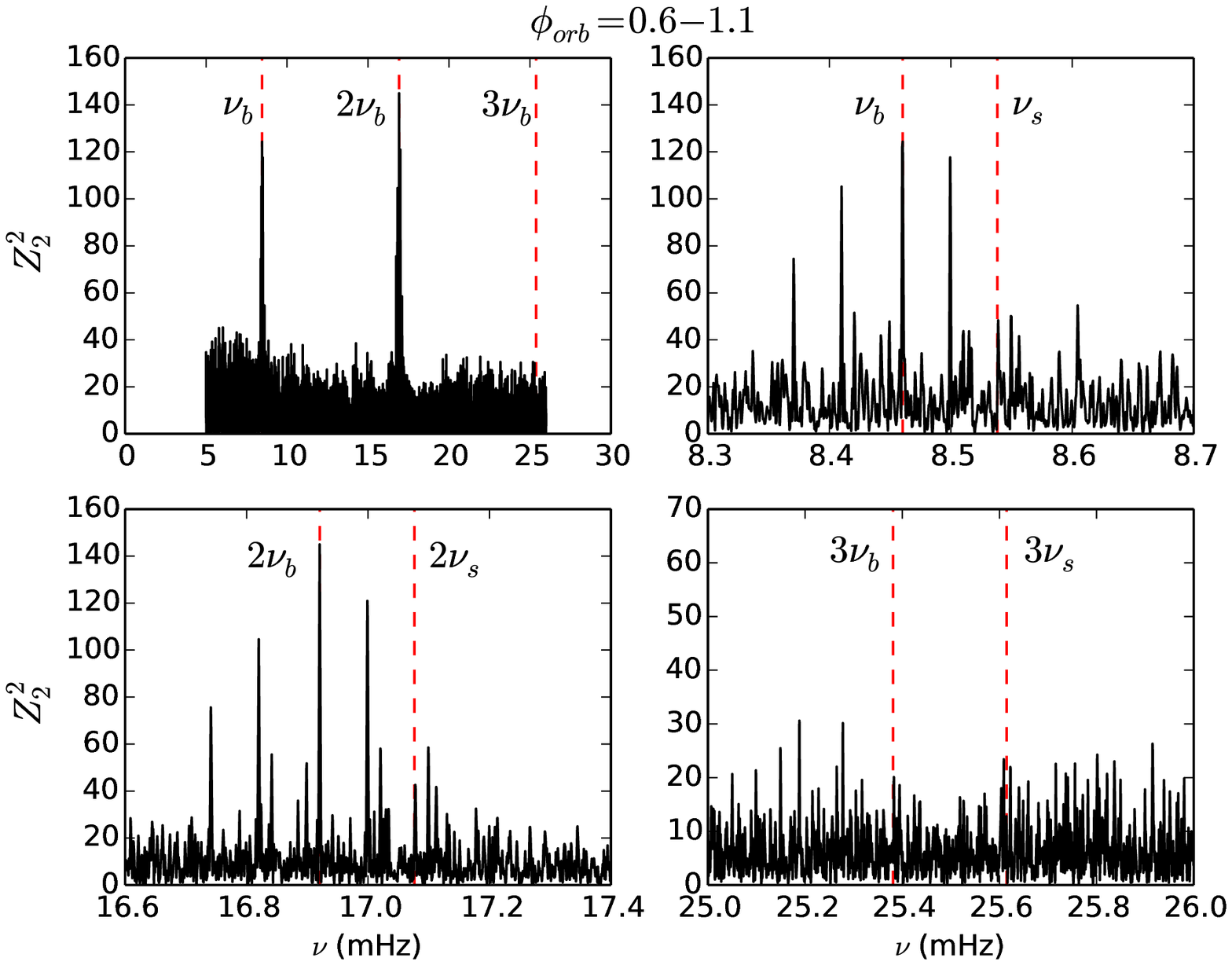}}
\caption{Orbital resolved $Z_2^2$ periodogram for Ni20  data set. $\nu_b$ and $\nu_s$ represents the                       
beat frequency and spin frequency of WD, respectively. The dotted lines in the left panel show the position of the average\
 between two frequencies.}
\label{nz}
\end{figure*}

In the dynamic pulse profile in Figure~\ref{map}, we can see that a migration of the pulse peak position at the orbital phase $\phi_o\sim 0.2-0.5$. This migration suggests an existence of a signal with another period.   In fact, the $Z^2_2$ periodogram for the Ni20 data  (Figure~\ref{z}) clearly indicates that a component of the X-ray emission from \src\ modulates with not only the beat frequency but also the WD's spin frequency. 
 Figure~\ref{spin} presents  dynamic pulse profile folded with the spin frequency (left panel) and integrated pulse profile (right panel) of the   Ni20 data.  In  the dynamic pulse profile,  we can see a relatively constant peak position in the spin phase during the orbital phase $\phi_{o}\sim0.2-0.5$. This suggests that  the signal of the WD's spin frequency appears in the X-ray data only at a specific orbital phase interval. 
As the right panel in Figure~\ref{spin} shows, 
the spin  pulse profile integrated over the orbital phase  can be described by  
 a single peak structure, which is different from the double-peak structure of the beat pulse profile.  

We divide the Ni20 data into two orbital intervals, $\phi_{o}=0.1-0.6$ and $\phi_{o}=0.6-1.1$, and 
perform the $Z^2_2$-test for each interval (Figure~\ref{nz}). The periodogram  of the orbital phase  $\phi_{o}=0.1-0.6$ shows  the signal of 
the spin frequency that is stronger than the signal of the beat frequency. During $\phi_{o}=0.1-0.6$, moreover, we can see a strong signal at the average of two frequencies (dotted vertical lines), and  the first harmonics of the averaged frequency has a power stronger than those of the $2\nu_b$ and $2\nu_{s}$ in the periodogram.  This supports the fact that the short-term modulation during $\phi_{o}=0.1-0.6$ is caused by the mixture between the beat frequency and spin frequency.   For $\phi_{o}=0.6-1.1$ (right panel in Figure~\ref{nz}), on the other hand, no signal of the spin frequency is found in the $Z_2^2$  periodogram, and the short-term modulation is dominated  by the beat frequency. 
We find that the fist harmonic of the beat frequency during the orbital phase  $\phi_{o}=0.6-1.1$ shows a $Z_2^2$ power similar to or stronger than the power of the fundamental frequency. This is consistent with the double-peak structure 
with a phase separation $\sim 0.5$  appeared in the pulse profile folded with the beat frequency. 

 We  note that the dynamic pulse profile of the X16 data  (left panel in Figure~\ref{map}) also shows a migration of the  peak position and therefore indicates an appearance of the spin frequency at  the orbital phase interval $\phi_{o}\sim 0.2-0.5$. The similar behavior of the  X16  and N20 data  suggests that the appearance of signal of the WD's spin frequency only  at a specific orbital phase interval is  regularly  repeated by every orbit. In section~\ref{discuss}, we discuss  the  mixture between the spin and beat frequencies in the short-term modulation and its dependency on the orbital phase.

\section{Spectral analysis}
\label{spana}
The spectrum of the X-ray emission from \src\  can be described by an optically thin thermal plasma 
emission with several different temperatures \citep{takata18}. In this study, we fit the generated spectra
 with a multi-temperature VMEKAL model \citep{mewe85,mewe86} using  \verb|Xspec| version 12.10.1f.  We apply  \verb|TBABS| model for modeling photoelectric absorption.

\begin{deluxetable}{lccc}
\tablecaption{Fittings for 2016 XMM-Newton data (0.2-12~keV bands) with three-temperature VMEKAL models. \label{x16spe}}
\tablewidth{0pt}
\tablehead{  
    & Takata et al. (2018) \tablenotemark{a}     & Model 1 \tablenotemark{a}     &  Model 2 \tablenotemark{b} }
\startdata
$N_{\rm H}$ ($10^{20}{\rm cm^{-2}}$) & $3.4^{+0.8}_{-0.8}$ & $1.2^{+0.3}_{-0.2}$  & $2.8^{+0.5}_{-0.5}$  \\
$F_e$ & $0.67^{+0.29}_{-0.21}$ & $0.45^{+0.07}_{-0.07}$& $0.28^{+0.06}_{-0.05}$\\
$kT_1$ (keV) & $8.0^{+2.8}_{-1.6}$ &$7.6^{+6.0}_{-1.9}$& $6.3^{+1.3}_{-1.0}$\\
$kT_2$ (keV) &$1.7^{+0.4}_{-0.3}$&$2.4^{+0.9}_{-0.7}$& $1.5^{+0.2}_{-0.2}$\\
$kT_3$ (keV) & $0.6^{+0.1}_{-0.1}$& $0.63^{+0.04}_{-0.05}$& $0.32^{+0.04}_{-0.03}$ \\
$\chi_{red}^2$ (dof) & 1.03 (404) &1.20 (406) & 1.08 (406)\\
\enddata
\vspace{-0.3cm}
\tablenotetext{a}{Except for iron, the abundances are fixed to the solar abundances [Anders and Grevesse (1989)]}
\tablenotetext{b}{The abundances of O, Ne, Mg, S and Si are linked to that of Fe. }
\end{deluxetable}

\subsection{Analysis for individual data set}
\begin{figure*}
\begin{tabular}{cc}
\begin{minipage}[t]{0.5\hsize}
        \centering
        \includegraphics[keepaspectratio,scale=0.45]{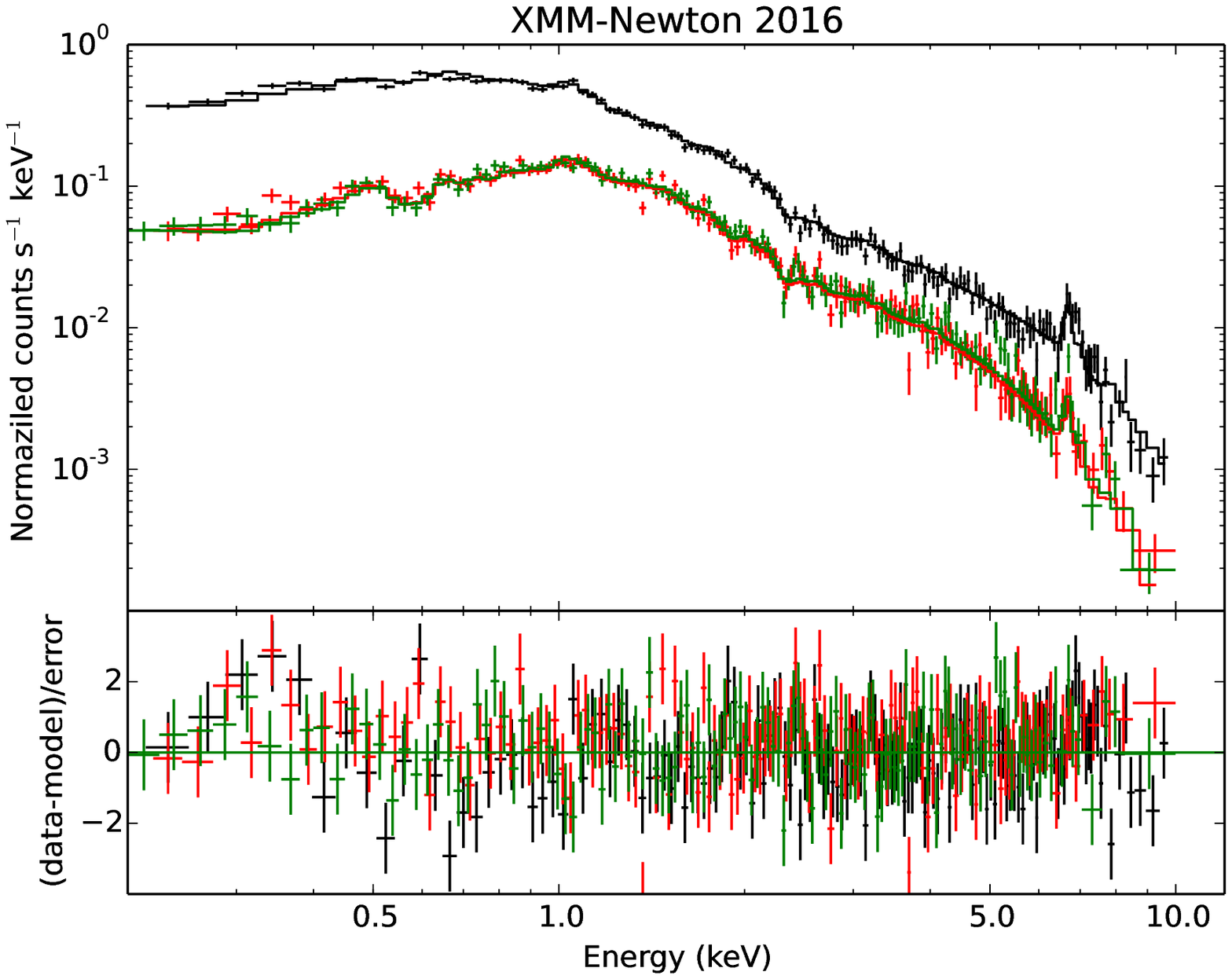}
\end{minipage} &
\begin{minipage}[t]{0.5\hsize}
        \centering
        \includegraphics[keepaspectratio,scale=0.45]{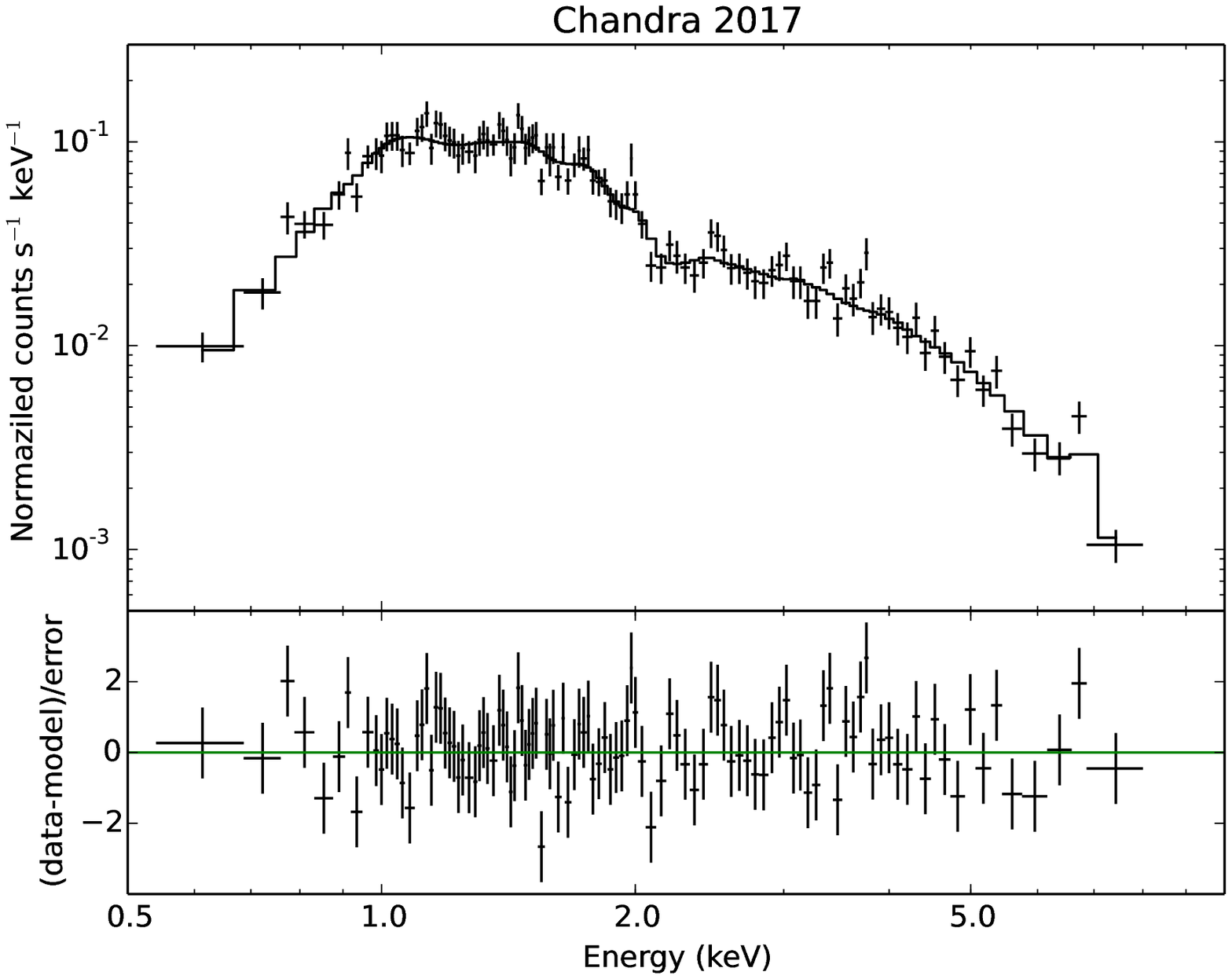}
\end{minipage} \\
\begin{minipage}[t]{0.5\hsize}
        \centering
        \includegraphics[keepaspectratio,scale=0.45]{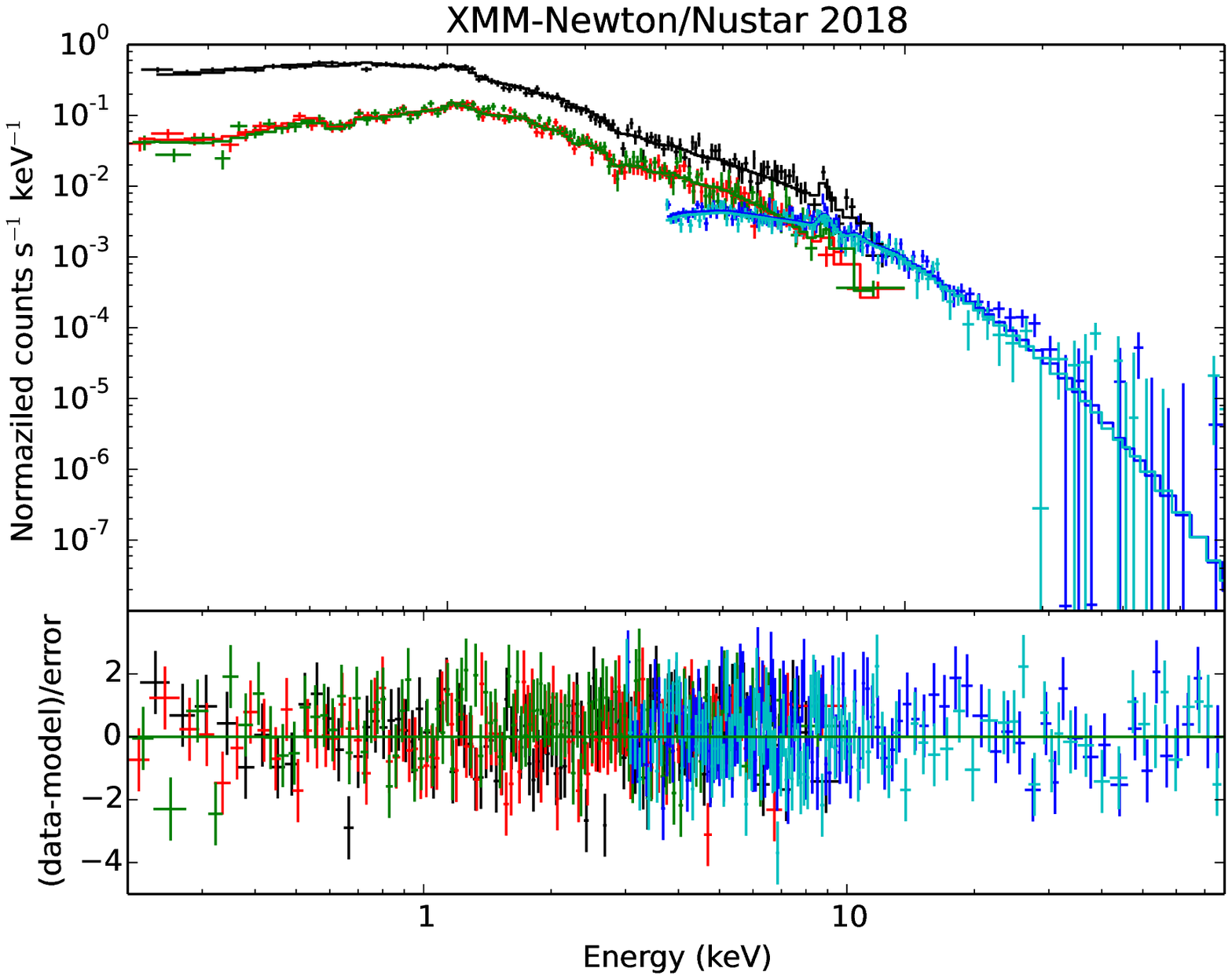}
\end{minipage} &
\begin{minipage}[t]{0.5\hsize}
        \centering
        \includegraphics[keepaspectratio,scale=0.45]{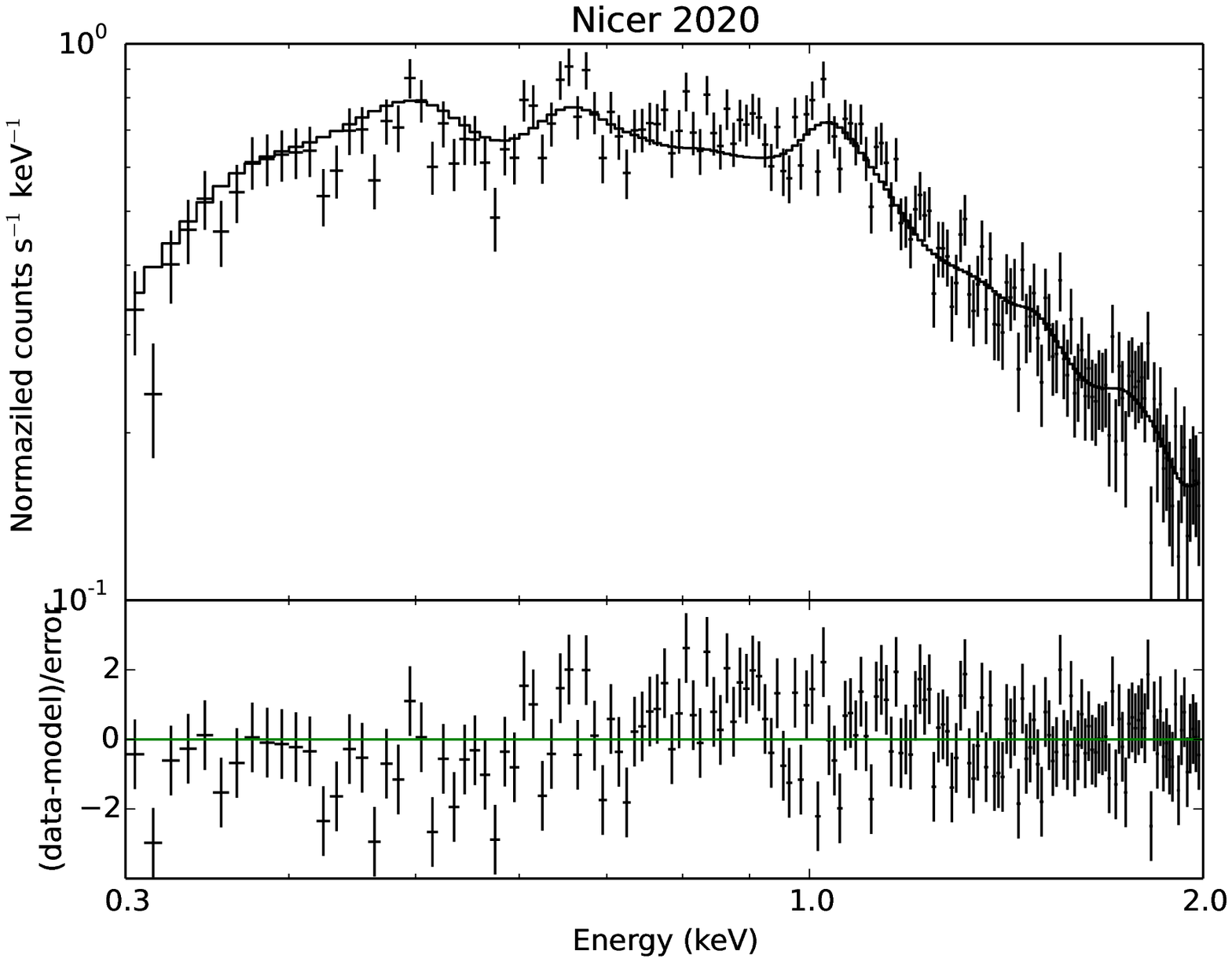}
\end{minipage}

\end{tabular}
\caption{X-ray spectrum of \src. Top left: 2016 XMM-Newton observation.  The black, red and green correspond to PN,MOS1 and 
MOS2, respectively. Top right: 2017 Chandra observation. Bottom left: 2018 XMM-Newton/NuSTAR observations. For XMM-Newton data, the black, red and green show the spectra of PN, MOS1 and MOS2, respectively. For NuSTAR data, the blue and cyan correspond to FPMA and FPMB data, respectively. Bottom right: 2020 NICER observation.}
\label{spectrum}
\end{figure*}

\subsubsection{Reanalysis for X16 data}
\label{rex16}
Although \cite{takata18} fit the spectrum of the X16 data with  a three-temperature VMEKAL model, 
  we reanalyze the data with the  updated calibration files of XMM-Newton observation. Table~\ref{x16spe} compares the best-fit parameters reported in \cite{takata18} with those in the new analysis. In \cite{takata18} (the second column), a reasonable fit is obtained by thawing  iron (Fe) abundance and by fixing other elements to the solar abundance.  In our new analysis, however, we find that the  reduced $\chi^2$
of the fitting with the  model (Model~1 in third column) becomes worse comparing to the previous result, and the best-fit parameters show a smaller hydrogen column density $(N_H$).

 To find a better fitting  model, we thaw each element step by step. We find that abundances
 of another five elements (O, Ne, Mg, S and Si) smaller than the solar abundance 
significantly improve the fitting results with a $F$ statistic value of 8.2, which means that the probability of this improvement being caused by chance is $2.3\times 10^{-7}$. Since the predicted abundances of
 those components are consistent with that of the iron within the errors,  Table~\ref{x16spe} (Model~2)
 shows the best-fit parameters  by linking the abundances between the iron and the five elements. 
We find that the hydrogen column density and two higher temperatures are consistent with previous result, but the abundance of the Fe and the lowest temperature decrease from the previous one.

\subsubsection{NICER data}
Since NICER has no imaging capability, we check whether   the spectra obtained with the process described in section~\ref{nicer}  is indeed  consistent with the X-ray emission from  \src. We  restrict the photon energy in the 0.3-2~keV band for the spectral analysis of the NICER data because of strong background contamination  in lower/higher energy bands.  We use the Model 2 of Table~\ref{x16spe}, in which   the abundances of O, Ne, Mg, S, Si and Fe elements are linked.  We find that the background subtracted spectrum is reasonably fitted by a two-temperature VMEKAL model with the best-fit hydrogen column density $N_H=4.6^{+0.8}_{-0.7}\times 10^{20}{\rm cm^{-2}}$,  temperatures $kT_1=2.6^{+0.4}_{-0.4}$~keV  and $kT_2=0.67^{+0.09}_{-0.06}$~keV ($\chi_{red}^2=1.15$ for 163 dof: third temperature component is not necessary). The Fe abundance of $0.27^{+0.11}_{-0.08}$ (solar abundance) is consistent with the  result of the X16 data in Table~\ref{x16spe}. A simple black-body radiation model or a power-law model cannot provide a reasonable fit. Hence, the generated spectrum  likely represents the X-ray emission from  \src.

The pulse profiles  folded with the beat frequency of the Ni20 data  shows a single-peak structure 
 in $\phi= {0.1-0.6}$ and a double-peak structure in $\phi={0.6-1.1}$ (Figure~\ref{ob}).  We therefore  create the orbitally resolved spectra  for $\phi_o={0.1-0.6}$ and $\phi_o={0.6-1.1}$, and we check any orbital evolution of the best-fit parameters with the two-temperature VMEKAL model. We do not find a significant difference in the best-fit parameters of  two orbital phase intervals.  This is not  unexpected because the pulsed emission is only $\sim$10\% of the total emission, and the spectrum is dominated by the un-pulsed component.

\subsubsection{Pile-up fraction of Chandra data}
\label{pileup}
First, we use the \verb|Web PIMMS| Version 4.11 to estimate pile-up fraction of the  Chandra data. Since the \verb|Web PIMMS| implements the APEC\footnote{http://atomdb.org/} model for the plasma emission, we fit the generated spectrum with a two-temperature APEC model with the \verb|Xspec|.  Since the data cannot constrain the hydrogen column density, we fix it at $N_H=3.0\times 10^{20}{\rm cm^{-2}}$. A reasonable fit is obtained with the  temperatures  $kT_1=5.1^{+1.2}_{-0.5}$~keV, $kT_2=1.0^{+0.4}_{-0.3}$~keV and the iron abundance $0.6^{+0.2}_{-0.3}$ ($\chi_{red}^2=1.0$ for 107 dof). With the frame time of the observation, $0.8$~s, the  \verb|Web PIMMS| estimates the pile-up fraction as $\sim 5$ percent. Second, we use the \verb|CIAO|'s fitting package, \verb|Sherpa|, to estimate the pile-up fraction. We fit the spectrum with a two-temperature VMEKAL components plus pile-up model \citep{davis01}. The fitting result shows the plasma temperatures $kT_1=4.1^{+0.5}_{-0.5}$~keV, $kT_2=0.8^{+0.3}_{-0.4}$~keV and abundance Fe$=0.67^{+0.14}_{-0.13}$. The estimated
 pile-up fraction is $\sim5$ percent (so called $\alpha$-parameter and $f$-parameter are $\sim 0.8$ and $0.89$, respectively), which is consistent with the result of the \verb|Web PIMMS|.

\subsubsection{Joint fitting for X18 and Nu18 data}
\begin{deluxetable}{lccc}
\tablecaption{Join fitting for the spectra (0.2-78 keV bands) of 2018 XMM-Newton and NuSTAR observations. \label{18spe}}
\tablewidth{0pt}
\tablehead{
    & Model~2     & Model 3 \tablenotemark{a}     &  Model 4 \tablenotemark{b} }
\startdata
$N_{\rm H}$ ($10^{20}{\rm cm^{-2}}$) & $4.0^{+2.1}_{-1.4}$ & $3.4^{+1.6}_{-1.1}$  & $2.6^{+0.6}_{-0.6}$  \\
$F_e$ & $0.22^{+0.05}_{-0.05}$ & $0.28^{+0.08}_{-0.08}$& $0.31^{+0.07}_{-0.06}$\\
$kT_1$ (keV) & $6.8^{+0.8}_{-0.6}$ &$4.1^{+0.4}_{-0.4}$&$2.7^{+0.3}_{-0.3}$\\
$kT_2$ (keV) &$1.5^{+0.2}_{-0.2}$&$0.82^{+0.07}_{-0.07}$&  $0.66^{+0.14}_{-0.09}$ \\
$kT_3$ (keV) & $0.34^{+0.32}_{-0.06}$& -& - \\
$\alpha$\tablenotemark{c}& - & $1.6^{+0.1}_{-0.3}$& $-0.20^{+0.5}_{-1.0}$\\
$E_c$\tablenotemark{d}(keV) &- &- & $4.2^{+1.9}_{-1.1}$\\ 
$\chi_{red}^2$ (dof) & 1.07 (546) &1.14 (546) & 1.07 (545)\\
\enddata
\vspace{-0.3cm}
\tablenotetext{a}{Two VMEKAL +  Power-law model.}
\tablenotetext{b}{Two VMEKAL +  Power-law plus exponential cut-off model.}
\tablenotetext{c}{Photon index of the power-law component.}
\tablenotetext{d}{Cut-off energy of the power-law plus cut-off component. }
\end{deluxetable}

The observations for the  X18 and Nu18 data  were carried out on 2018 February 18 and 19, respectively. This joint observation enables us to investigate X-ray spectrum of \src\ over a wide energy range in the 0.2-78keV bands. In particular, we can investigate the behavior of the X-ray spectrum above 10keV bands. We perform a joint fit for the two spectra by introducing a  \verb|constant| factor to account for cross-calibration mismatch between the XMM-Newton and NuSTAR observations. During the fitting process, we fix the \verb|constant| factor for the X18 data at unity,  and obtain a  constant factor of   $\sim 1.1$ for the  Nu18 data.

First we fit the spectrum with a three-temperature VMEKAL model (Model 2 in Table~\ref{18spe}), in which we 
link the abundances of six elements (O, Ne, Mg, S, Si and Fe) and obtain the best-fit parameters as $N_H=4.0^{+2.1}_{-1.4}\times 10^{20}{\rm cm^{-2}}$, $kT_1=6.8^{+0.8}_{-0.6}$~keV, $kT_2=1.5^{+0.2}_{-0.2}$~keV, $kT_3=0.34^{+0.32}_{-0.06}$ and $F_e=0.22^{+0.05}_{-0.05}$ with $\chi_{red}^2=1.07$ for 546~dof., which are consistent with the results of X16 data within the errors (Table~\ref{x16spe}).

To examine the spectral behavior above 10keV bands, we replace one VMEKAL component to a power-law model, and fit the spectrum with  \verb|constant*tbabs*(vmekal+vmekal+powerlaw)| (Model 3 in Table~\ref{18spe}). We obtain the best-fitting photon index of $\alpha=1.6^{+0.1}_{-0.3}$,  
but we find that   the $\chi_{red}^2\sim 1.14$ of the fitting becomes worse in comparison  to the three-temperature VMEKAL model. This would suggest that the observed spectrum does not extend in the hard X-ray band  with a single power-law form.

Finally we  consider possibility of a spectral cut-off feature  by  replacing  the power-law component to a power-law plus an exponential cut-off model. We  fit the data  with \verb|constant*tbabs*(vmekal+vmekal+cutoffpl)| (Model 4 in Table~\ref{18spe}), and obtain an acceptable fit with $\chi_{red}^2\sim 1.07$ for 545 dof, which is comparable to that of three VMEKAL model. The best-fit parameters, however, show  a  very hard photon index of $\alpha\sim-0.20^{+0.5}_{-1.0}$ with a cut-off energy $E_c\sim 4.2^{+1.9}_{-1.1}$~keV, and indicate that the spectrum of the third component in higher energy band has a convex shape. This spectral property is consistent with a thermal emission rather than a non-thermal emission, and therefore the optically thin thermal plasma emissions with three different temperatures describes better  the X-ray emissions from \src\ over the 0.2-78keV bands.

\subsection{Investigation for long-term evolution}
\begin{deluxetable}{lcccc}
\tablecaption{Simultaneous fittings for the spectra (0.2-78 keV bands) with three-temperature VMEKAL models and by linking three temperatures.\label{vmekal1}}
\tablewidth{0pt}
\tablehead{
    & X16              &Ch17     & X18/Nu18              & Ni20 }
\startdata
$N_{\rm H}$ ($10^{20}{\rm cm^{-2}}$) &  \multicolumn{4}{c}{$3.8^{+0.4}_{-0.4}$}  \\
$F_e$ & \multicolumn{4}{c}{$0.25^{+0.03}_{-0.03}$} \\
$kT_1$ (keV) &  \multicolumn{4}{c}{$6.4^{+0.6}_{-0.4}$} \\
$kT_2$ (keV) & \multicolumn{4}{c}{$1.4^{+0.19}_{-0.05}$} \\
$kT_3$ (keV) & \multicolumn{4}{c}{$0.30^{+0.02}_{-0.02}$}   \\
$N_1$\tablenotemark{a} $(10^{-4})$ & $12.7^{+0.5}_{-1.0}$& $13.1^{+1.0}_{-1.0}$ &$11.0^{+0.6}_{-1.1}$ &\
$8.4^{+0.8}_{-0.9}$  \\
$N_2$\tablenotemark{a} $(10^{-4})$ & $5.0^{+1.3}_{-0.7}$ &  $6.9^{+1.6}_{-1.5}$
&$5.3^{+1.3}_{-0.8}$  & $4.6^{+0.9}_{-0.8}$ \\

$N_3$\tablenotemark{a} $(10^{-4})$ & $2.5^{+0.4}_{-0.4}$&  $1.6^{+1.4}_{-1.4}$
 & $1.6^{+0.4}_{-0.4}$ &$0.8^{+0.2}_{-0.2}$ \\

$F_{0.3-10}$\tablenotemark{b} &  $2.88^{+0.03}_{-0.03}$  &$3.05^{+0.09}_{-0.09}$
&$2.54^{+0.04}_{-0.04}$ & $1.97^{+0.03}_{-0.03}$     \\
$\chi_{red}^2$ (dof) & \multicolumn{4}{c}{1.12 (1232)}  \\
\enddata
\vspace{-0.3cm}
\tablenotetext{a} {Normalization of the VMEKAL component in units of $10^{-14}/(4\pi d^2)\int n_en_HdV$\
, where $d$ (cm) is the distance to the source.}
\tablenotetext{b}{Unabsorbed flux in the 0.3-10~keV band in units of $10^{-12}~{\rm erg~s^{-1}cm^{-2}}$\
 .}
\end{deluxetable}

\begin{deluxetable}{lcccc}
\tablewidth{0pt}
\tablecaption{Simultaneous fitting for the spectra (0.2-78 keV bands) with three-temperature VMEKAL models and by unlinking lower two
temperatures. \label{vmekal2} }
\tablewidth{0pt}
\tablehead{
    & X16        & Ch17            & X18/Nu18              & Ni20
}
$N_{\rm H}$ ($10^{20}{\rm cm^{-2}}$) &  \multicolumn{4}{c}{$3.6^{+0.4}_{-0.3}$}  \\
$F_e$& \multicolumn{4}{c}{$0.26^{+0.03}_{-0.03}$} \\
$kT_1$ (keV) &  \multicolumn{4}{c}{$6.8^{+0.6}_{-0.5}$}  \\

$kT_2$ (keV) & $1.6^{+0.1}_{-0.2}$ & $1.6^{+0.6}_{-0.3}$
&$1.6^{+0.2}_{-0.2}$&$1.6^{+0.3}_{-0.2}$ \\

$kT_3$ (keV) & $0.29^{+0.02}_{-0.02}$ & $0.3^{+0.7}_{-0.2}$
&$0.32^{+0.07}_{-0.04}$ & $0.61^{+0..08}_{-0.11}$   \\

$N_1$   $(10^{-4})$ & $11.8^{+1.0}_{-1.0}$&$11.9^{+1.6}_{-2.5}$
&$10.1^{+1.1}_{-1.1}$ &$7.6^{+1.6}_{-2.0}$  \\

$N_2$ $(10^{-4})$ & $6.0^{+1.1}_{-1.1}$ &$8.0^{+2.6}_{-2.2}$
&$6.2^{+1.3}_{-1.3}$  & $4.9^{+1.9}_{-1.6}$ \\

$N_3$ $(10^{-4})$ & $2.2^{+0.4}_{-0.3}$& $1.7^{+100.0}_{-1.4}$
  & $1.4^{+0.4}_{-0.3}$ &$1.0^{+0.3}_{-0.2}$ \\

$F_{0.3-10}$ & $2.86^{+0.03}_{-0.03}$ &$3.05^{+0.09}_{-0.09}$
 &$2.51^{+0.04}_{-0.04}$ & $1.93^{+0.04}_{-0.04}$   \\

$\chi_{red}^2$ (dof) & \multicolumn{4}{c}{1.09 (1226)} \\
\enddata
\vspace{-0.3cm}
\end{deluxetable}

\begin{deluxetable*}{lcccc}[b]
\tablecaption{Fittings for 2016 and 2018 XMM-Newton data (0.2-12keV bands) with three-temperature MEKAL models.\label{vmekal3}}
\tablewidth{0pt}
\tablehead{
    &  \multicolumn{2}{c}{2016} &\multicolumn{2}{c}{2018} \\ 
    & 0-0.35/0.85-1             & 0.35-0.85      &   0-0.35/0.85-1            & 0.35-0.85                
}
\startdata
$N_{\rm H}$ ($10^{20}{\rm cm^{-2}}$) &  \multicolumn{4}{c}{$2.9^{+0.5}_{-0.4}$}  \\

$F_e$\tablenotemark{b} & \multicolumn{4}{c}{$0.27^{+0.05}_{-0.04}$} \\

$kT_1$ (keV) &  \multicolumn{4}{c}{$7.0^{+1.5}_{-1.2}$} \\

$kT_2$ (keV) &  \multicolumn{4}{c}{$1.6^{+0.1}_{-0.2}$} \\

$kT_3$ (keV) & \multicolumn{4}{c}{$0.32^{+0.04}_{-0.02}$}\\

$N_1$ $(10^{-4})$ & $10.3^{+1.5}_{-1.2}$&$12.3^{+1.7}_{-1.4}$
&$9.9^{+1.4}_{-1.2}$ &$10.4^{+1.7}_{-1.4}$  \\

$N_2$\tablenotemark{a} $(10^{-4})$ & $5.6^{+1.6}_{-1.9}$ & $6.5^{+1.8}_{-2.1}$ 
&$5.1^{+1.6}_{-1.7}$  & $6.7^{+1.9}_{-3.0}$ \\

$N_3$\tablenotemark{a} $(10^{-4})$ & $1.4^{+0.4}_{-0.3}$ &$2.2^{+0.5}_{-0.4}$
 & $1.2^{+0.4}_{-0.4}$ &$1.1^{+0.5}_{-0.5} $ \\
 
$F_{0.3-10}$\tablenotemark{c} & $2.53^{+0.04}_{-0.04}$  &$3.03^{+0.04}_{-0.04}$ &$2.38^{+0.05}_{-0.05}$ & $2.62^{+0.06}_{-0.06}$   \\
$\chi_{red}^2$ (dof) & \multicolumn{4}{c}{0.99 (1229)} \\ 
\enddata
\vspace{-0.3cm}
\end{deluxetable*}
We fit all spectra of X16, Ch17,  X18/Nu18 and Ni20, simultaneously, with a three-temperature VMEKAL model  by linking  the hydrogen column densities in  different data sets.   We apply the pile-up model implemented in the \verb|Xspec|, and fit all spectra with  the model \verb|pileup*constant*tbabs*(vmekal+vmekal+vmekal)|, for which  we turn on the pile-up model  only for the Chandra data.  Since the pile-up  parameters $\alpha$ and $f$ are not constrained by  the current data, we  fix the those parameters to be the best-fit values $0.8$ and  $0.89$, respectively, obtained in section~\ref{pileup}. The  \verb|constant| factors for the X16, Ch17, X18 and Ni20  data  are fixed to unity. 
 We link the  abundances of the six elements (O, Ne, Mg, S, Si and Fe) and fix other elements to the solar abundances, as discussed in section~\ref{rex16}. A  long-term evolution of the spectral properties is discussed 
with following three  cases.

First, by assuming a constant plasma temperature and  flux with time, we link the three temperatures and 
 normalization factors for all data sets. We find that the fitting model  is unacceptable 
 with a reduced chi-square of  $\chi_{red}^2\sim 2.3$ for 1232 dof. Then we unlink the normalization factors, but keeping the linking for the temperatures. We obtain a reasonable fit with $\chi_{red}^2\sim 1.12$ for 1216 dof. Table~\ref{vmekal1} and Figure~\ref{spectrum} summarize  the best-fit parameters and spectra, respectively.  We find that the best-fit parameters  are consistent with the results of the fitting  for the X16 data only (Model 2 in Table~\ref{x16spe}).  In Table~\ref{vmekal1}, we can see that the 0.3-10~keV unabsorbed flux  deviates from a constant flux with time.  One possibility is an intrinsic temporal evolution of the X-ray emission of \src\, as indicated  by the evolution of the pulse profile discussed in section~\ref{timing}. For example, we can see in Table~\ref{vmekal1} that the normalization factors ($N_1$, $N_2$ and $N_3$) of the Ni20 data are smaller than the values of the X16 data, indicating the volume of the emission region evolves with time. 

 We also unlink the abundances for the different data sets to see how the best-fit parameters evolve. We obtain the best-fit abundances as $F_e=0.21\pm 0.04$ for X16, $0.33^{+0.22}_{-0.15}$ for Ch17, $0.22\pm0.04$ for X18/Nu18 and $0.54^{\pm 0.18}_{-0.13}$ for Ni20, respectively. The abundance of Ni20 data is higher than those of the other data sets. This may be due to the 
fact that the spectrum of Ni20  is limited  in the 0.3-2.0~keV bands, which does not cover the 6.8~keV line emission from He-like $Fe$, and hence the abundance of Fe may be overestimated. We can see that other best-fit parameters are consistent with the results in Table~\ref{vmekal1} within the errors.

Second, to investigate another  possibility that the change of the X-ray flux is associated with a change of the temperature of the emitting plasmas, we re-fit the data by unlinking the lower two temperature components ($kT_2$ and $kT_3$). For the highest temperature,  since the Ch17 and Ni20 data sets cannot constrain the value, we keep the linking. An improvement  of the fitting  is found with a $F$ static value 5.5, which means that the probability of this improvement being caused by chance is $\sim10^{-5}$. Table~\ref{vmekal2} summarizes the best-fit parameters of this model. The X16, Ch17 and X18/Nu18 data are consistent with a constant temperature, although the error range for $kT_2$ of Ch17  is large. We find  however that the lowest  temperature $kT_3\sim 0.61$~keV of the Ni20 data  is  higher than the values $kT_3\sim 0.3$~keV of the other data sets, suggesting the plasma temperature shows a temporal evolution.  
We find that this tendency of the temperatures does not change, even though we unlink  the  abundances for the different data sets. In such a case,
 we obtain the abundances of  $F_e=0.25\pm 0.05$ for X16, $0.50^{+0.28}_{-0.23}$ for Ch17, $0.23\pm 0.05$ for X18/Nu18, and $0.45^{+0.18}_{-0.12}$ for Ni20, respectively.

Finally,  we can see in Figure~\ref{exposure} that the different observations were carried out with a different exposure distribution  over the orbital phase. Since the observed X-ray emission shows an orbital modulation with a large amplitude, it is possible that the different coverage over the orbit produces an apparent long-term evolution of the flux. To investigate such a  possibility,  we check  orbital phase-resolved spectra for the X16 and X18 data.  As we can see in Figure~\ref{exposure}, 
the exposure distribution at the orbital phase $\phi_{o}\sim 0.35-0.85$ is relatively uniform for both X16 and X18 data sets. Hence, we divide the orbital phase into two intervals, $0-0.35/0.85-1$ and $0.35-0.85$, and obtain a  spectrum for each orbital interval. 

Table~\ref{vmekal3} summarizes the best-fit parameters  with a three-temperature VMEKAL model; we  link the temperatures of the two data sets and the abundances for 6 elements (O, Ne, Mg, S, Si and Fe).  In each observation, we find that the flux in $\phi_{o}=0.35-0.85$ is higher than that in  $\phi_{o}=0-0.35/0.85-1$, and the flux averaged  over the two phase intervals is consistent with the flux obtained from the whole data of each observation (Table~\ref{vmekal1}). In each orbital phase interval, the flux of the X16 data is higher than that of the X18 data. These results  suggest that the difference in the exposure distribution over the orbital phase  is not the main reason of the difference between the fluxes of the X16 and X18 data sets.  Although more concrete evidence is required with more data,  the non-pulsed X-ray emission, as well as the pulsed emission,  of \src\ likely experiences a long-term evolution on a timescale of years. 

\subsection{Pulsed spectrum}
\label{pulspe}

\begin{figure}
 \epsscale{1.}
\centerline{
\includegraphics[scale=0.5]{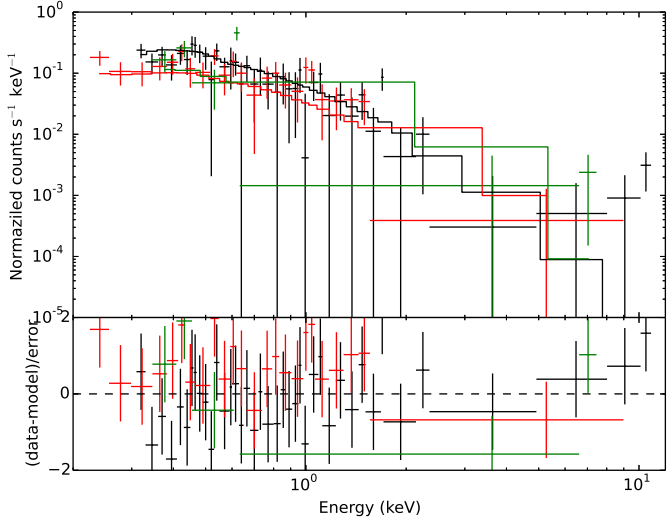}}
\caption{Spectrum of the pulsed component. The red, green and black point/model show the X16, X18 and Ni20 data sets, 
respectively. }
\label{puse-spe}
\end{figure}

\cite{takata18} examine the spectrum of the pulsed component with the X16 data and find evidence of a  power-law component with a photon index $\Gamma=2.3\pm0.5$. In the new analysis, we include  the X18 and Ni20 data set, but exclude MOS1/MOS2 data of the X16/X18 data because of a small amount of the  photon counts. 
 We define the on-pulse phase and off-pulse phase based on  the pulse profiles folded with the beat frequency, and  we generate the spectrum of the pulsed component by subtracting the off-pulse spectrum from the on-pulse spectrum.  As  indicated in Figure~\ref{light},  we define the on-pulse phase to cover the main peak for the X16/18 data, and  to cover main/secondary peaks for the Ni20 data.  We generate on-pulse and off-pulse spectra using the \verb|XMMSAS| for the X16/X18 data and using  the \verb|Xselect| for the N20 data. We  use the \verb|grppha| task to group the spectral bins so that the count in each bin after subtracting the off-pulse emission remains $>$40 photons for the  X16/X18 data  and $>100$ photons for the Ni20 data.

During the fitting, we fix the hydrogen column density at $N_H=3.5\times 10^{20}~{\rm cm^{-2}}$ inferred from the phase averaged spectrum. The power-law model provides a reasonable fit with  a photon index of $\Gamma=2.80^{+0.28}_{-0.25}$ ($\chi_{red}^2=0.8$) and an  unabsorbed flux of  $F_{0.2-2~keV}=2.37^{+0.29}_{-0.28}\times 10^{-13}~{\rm erg~cm^{-2}s^{-1}}$. Such a large photon index  is  consistent with a slope expected from  the ratio of the fluxes in optical and X-ray bands.    An optically thin plasma emission model (e.g. two-temperature MEKAL model) also provides a comparable goodness of the fit with temperatures $kT_1=1.8^{+1.5}_{-1.3}$~keV and $kT_2=0.11^{+0.03}_{-0.08}$~keV, but we cannot constrain the abundance.  The black-body radiation model predicts an  emission size of $\sim 0.1$~km with a source distance $d\sim 0.1$~kpc, which is unreasonably small. 

 \cite{garnavich19} find that the photon index of the optical emission varies significantly with the orbital phase and with the emission magnitude. With the current X-ray data quality, however, it is difficult to investigate such a dependency  of the properties of the pulsed emission on the beat phase or on the orbital phase.

As discussed in section~\ref{timing}, we cannot find the beat  signal in the $>2$~keV band.  If the spectrum of the pulsed emission could extend beyond 2~keV band with a power-law distribution, the expected flux in the 2-10~keV band is $F_{2-10~keV}=3.2^{+1.2}_{-0.8}\times 10^{-14}~{\rm erg~cm^{-2}s^{-1}}$, which is $\sim2$\% of the total emission in the 2-10~keV  band. Using all data sets,  we have a  total of  $\sim$55,000  events in the $>$2~keV band. Since the background emission dominates in  the  $>$2~keV band of the Ni20 data, the events related to the pulsed emission will be much less than  $\sim$1,100. With current collected  photon counts, therefore, the  significance of the pulsation in the $>2$~keV band will be too small to be detected (i.e., $<5\sigma$), even if the power-law component extends to that energy band. 
   
\section{Sumary and Discussion}
\label{discuss}
We have studied the X-ray emission from \src\ with the archival data taken in 2016-2020. In the short-term modulation,  the periodogram using the  Ni20 data clearly indicates a modulation of the spin frequency of WD,  indicating  that part of the X-ray emission is coming from the WD's magnetosphere. The  short term modulation  during the orbital phase  $\phi_{o}\sim 0.1-0.6$ is caused by the mixture between the beat frequency and the spin frequency, while the intensity peak during $\phi_{o}\sim 0.6-1.1$ is regularly repeated with the beat frequency. The X16 data also indicates such  a  modulation behavior over the orbital phase. Hence, the appearance of the WD's spin frequency in the X-ray emission  only at the specific orbital  
phase is  repeated by every orbit. In the  long-term variation,  we find that the  observed flux level in 2016/2017 was  higher than those  measured in 2018/2020, and the amplitude of the orbital waveform of the  2020 NICER data might be larger than those of other data sets taken in 2016-2018.  Moreover, the beat pulse profile averaged over the orbital phase  changed from a single-peak structure in 2016/2018 to  a double-peak structure in 2020. These results suggest that  the X-ray emission from \src\ experiences a long-term evolution on a timescale of years.   The spectrum of the pulsed emission  can be fit by a power-law function with a photon index $\Gamma\sim 0.25-0.31$.

A mixture of the beat frequency and the spin frequency in the short-term modulation is also observed in the optical emission from \src\ \citep{marsh16,potter18a,potter18b}. According to the figure~3  in \cite{potter18b} (the periodogram  of the optical data), the beat signal is much stronger than the spin signal in the optical band.  In the X-ray bands,  the signal of the  WD's spin frequency during the orbital phase $\phi_{o}\sim 0.1-0.6$ is similar strength to or may be stronger than that of  the beat frequency, while the signal of the spin frequency disappears during the orbital phase $\phi_{o}\sim 0.6-1,1$, as we discussed in section~\ref{timing}.  This shows
 that  a component of the X-ray emission  modulating  with the WD spin frequency appears and 
disappears in the observation over the orbital phase. We may interpret the signal of the beat frequency and the spin frequency in the X-ray emission  as  a result of the synchrotron emission from two kind of the electron populations characterized by a larger and a smaller pitch angles, respectively.
 A brief discussion of this interpretation is as follows. 

We assume a non-thermal emission process for the observed X-ray emission modulating with the spin and/or beat frequencies.  
The non-thermal process of \src\ has been discussed  that an interaction between the magnetosphere of the WD  and companion star  may form a bow shock or cause a magnetic dissipation process, which could accelerate the electrons from the companion star \citep{geng16, buckley17, takata17, singh20}. \cite{garnavich19} and \cite{lyutikov20}, on the other hand,  propose that it is related to the magnetic reconnection event  in the magnetosphere of the WD because of  the interaction with the flow from the companion star. \cite{bednarek18} proposes a hadronic model, in which  the electrons and  protons  are accelerated in a strongly magnetized turbulent region around the M-type star,  and  the primary electrons and/or secondary electron/positron pairs produce a non-thermal emission from \src.  We assume that  the accelerated electrons at the interaction region are trapped in the WD magnetosphere and, produce the non-thermal emission through the synchrotron radiation. 

The  modulation of the beat frequency has been interpreted that the energy and/or particle injection for the synchrotron radiation is enhanced  when the magnetic pole points toward  the companion star \citep{geng16,potter18b}. \cite{takata17} suggest on the other hand  that the synchrotron  emission around the magnetic mirror  points, where the pitch angle of the electrons becomes $90$ degree,  produces a pulsed emission. In their model, if the WD's magnetic pole is inclined from  the spin axis, the electrons injected into the WD's magnetosphere at different spin/beat  phases  migrate to the first magnetic mirror points with different travel times. With the  effect of the travel times,  the emission from  different magnetic mirror points can enhance the observed intensity at certain spin/beat phase, and results in the formation of the pulse. This model expects  that the pulse profile evolves over  the orbital phase,  because the position of the  WD relative to the companion star  measured from the Earth depends on the orbital phase.

In the observed X-ray emisison, the modulation with the WD's spin frequency appears only at a specific orbital phase interval. \cite{potter18b} discuss synchrotron radiation from a region fixed in the WD's magnetosphere to explain the optical pulsed emission from \src. They demonstrate that by choosing an appropriate pitch angle of the synchrotron radiation,  the pulsed emission  modulating with the spin frequency appears brighter during the orbital phase $\phi_o\sim0.2-0.45$ only, because of a beaming effect of the synchrotron radiation.  Our speculative interpretation therefore is that the pulsed X-ray emission of \src\ are produced by  two kinds of the electron populations; electrons  having a larger pitch angle and one having a smaller pitch angle.  The emission that modulates with the beat frequency is produced by the population  with a larger pitch angle, while the emission that modulates with the spin frequency is produced by the population  with a smaller pitch angle.  

The electrons  that are injected from the interaction region will initially move toward the WD surface along the magnetic filed line.  Because of the conservation of the  first adiabatic invariance, the pitch angle increases as the electrons approach to the  WD surface, because the strength of the magnetic field increases.   If the electrons with a pitch angle $\theta_0$  are injected into the WD magnetosphere at a radius distance $a$ from the WD,  the pitch angle will evolve with the  radial distance, $r$, measured from the WD as  
\begin{equation}
\sin\theta(r)=\left(\frac{a}{r}\right)^{3/2}\sin\theta_0,
\label{pitch}
\end{equation}
where we assume a magnetic dipole field and ignore a change of the Lorentz factor due to  the synchrotron radiation loss. The electrons injected with a larger pitch angle will lose their energy at or near the magnetic mirror point.  Since the emission 
from the electrons with a larger pitch angle covers a wider area in the sky,  it can be observable over the orbital phase.   We expect that such an emission modulates with the beat frequency.

For the electrons injected with a smaller pitch angle, on the other hand, they can move in the  deeper WD's magnetosphere while  keeping a small pitch angle. If the magnetic field is strong enough, the electrons  may  lose their energy through the synchrotron radiation before reaching the magnetic mirror points.  A simple quantitative estimation is the following. As an electron moves along the dipole magnetic field line toward the WD, the synchrotron loss timescale ($\tau_{syn}$) decreases since the pitch angle and magnetic field strength increase. When the synchrotron loss timescale becomes of the order of the dynamic timescale, $\tau_{c}=r/c$, the electrons significantly lose the energy. By equating $\tau_{syn}=\tau_c$, and using the equation~(\ref{pitch}), 
we estimate the Lorentz factor of the  electron  (i) that are injected with a pitch angle $\theta_0$ at $r=a$ and  (ii) that  emit the typical synchrotron photon energy $E_{syn}$ as
\begin{equation}
\gamma=\frac{2}{3}\left[\frac{16\pi}{27}r_e^{9/16}r_g^{-1/8}R_{WD}^{3/8}a^{3/16}\left(\frac{E_{syn}}{hc}\right)\right]^{16/23}\sin^{2/23}\theta_0,
\end{equation}
where $r_e=e^2/m_ec^2$ and $r_g=m_ec^2R_{WD}^3/(e\mu_{WD})$ with $\mu_{WD}$ and $R_{WD}$ being the dipole moment and  radius 
of the  WD, respectively. The radial distance, $r_s$, at the emission region and the pitch angle at $r_s$ become
\begin{equation}
r_s=\left[\frac{16\pi}{27}r_e^2r_g^{-3}R_{WD}^9a^{9/2}\left(\frac{E_{syn}}{hc}\right)\right]^{2/23}\sin^{6/23}\theta_0,
\label{rs}
\end{equation}
and
\begin{equation}
\sin\theta(r_s)=\left[\frac{16\pi}{27}r_e^2r_g^{-3}R_{WD}^9a^{-7}\left(\frac{E_{syn}}{hc}\right)\right]^{-3/23}\sin^{14/23}\theta_0,
\end{equation}
respectively.

We may apply  $\mu_{WD}=5\times 10^{34}~\rm{G~cm^{3}}$ and $R_{WD}=10^9$~cm for the WD parameters of \src\ \citep{buckley17},  and $a=8\times 10^8$~cm of the separation between centers of two stars  as the injection distance (since the companion star, whose radius is   $\sim 2\times 10^{10}$~cm,  occupies a large portion of the space between centers of the two stars, this distance of the  injection may be  a  crude approximation).
 Then, if the initial pitch angle is very small, for example  $\sin\theta_0=0.01$, 
we obtain  $\gamma\sim 9000$, $\sin\theta(r_s)\sim 0.06$ and $r_s= 2\times 10^{10}$cm, respectively,  for the photon energy $E_{syn}=1$~keV. Such an electron therefore  loses the energy while keeping the small pitch angle. The emission from the small pitch angle approximately directs along the magnetic field line, and it is observable by the observer whose line of sight is parallel to the direction of the magnetic field  \textit{at the radial distance $r_s$}. We expect that the  X-ray emission from such a small pitch angle is observed  only at a specific $spin~phase$ interval (like the  pulsar  emission) and hence the observed emission modulates with  the spin frequency. 

The detectability of the emission from a small pitch angle for \src\ will depend on the orbital phase because of the beaming effect. We assume that new high-energy electrons on a WD's magnetic field line are injected when the magnetic field line interacts with the companion star magnetosphere/outflow. Since the dynamical timescale ($\sim$ one second)  and  the synchrotron timescale at the emission point  are much shorter than the WD's spin period, we anticipate that, in terms of the spin phase,  the ejected electrons  immediately lose most of  their energy after the injection.  At a specific orbital phase interval, therefore, if there is no electron injection when the magnetic field at the distance $r_s$ directs toward the observer, we cannot observe the X-ray emission modulating with the WD's spin.  This beaming effect  could explain why the signal of the WD's spin frequency  in the observed X-ray emission appears and disappears over the orbital phase.

Finally, another unique property is that \src\ shows a long-term evolution of the X-ray emission. In particular, the X-ray light curve folded with the beat frequency clearly indicates  an evolution of the pulsed component on a time-scale of years. This evolution could be caused either by an fluctuation or secular  evolution of the state of the system.  
The change of the pulse profile and orbital modulation have been discovered in the X-ray emission from  an intermediate polar (IP), XY Arietis \citep{norton07}, which is the   accreting  magnetic white dwarf, and the observed X-ray emission mainly  originates from  the white dwarf surface hit by the accretion flow.   XY Arietis experienced an outburst because of a  substantial increase of the accretion rate in 1996 \citep{hellier97}, and  the increase of the mass transfer  was probably caused by  a result of  a thermal-viscous disk instability \citep{hameury17} or an enhancement mass transfer rate from the companion star.  \cite{hellier97} report that the spin pulse profile evolved from a  double-peaked profile to a single-peaked profile during the outburst.  \cite{norton07} find that the spin pulse profile and the orbital waveform of XY Arietis also show a long-term evolution in the quiescent state.  They interpret the  long-term evolution in quiescent state with a precession model of an accretion disk.

\src\ is a  different type of  WD's binary system from the IP in the sense that there  is no  accretion disk in the system and the pulsed X-ray emission is likely produced by the synchrotron radiation of the relativistic electrons in the WD's magnetosphere.  It has been expected that the modulation peak in the beat phase occurs   when the magnetic poles direct toward the companion star. The double peak structure of the Ni20 data (and optical data) with a phase separation $\sim 0.5$ implies the emission from the northern and southern hemispheres. The single-peak structure observed in the  X16 and X18 data may indicate that  the X-ray emission from one hemisphere in 2016/2018 was intrinsically dim because of an  anisotropic energy injection from the interaction region. If this interpretation is true, the emerging secondary peak in the Ni20 data might be due to  an increase of the energy injection to the hemisphere. 

 Another possible interpretation can be that the X-ray  emission from one hemisphere was geometrically  invisible in 2016/2018 observations, and it became visible  in 2020. This interpretation could be possible if the spin axis of the WD is precessing.  
\cite{schwarzenberg92} estimate the precession period would be $\sim10^4-10^5$ times the WD's spin period, which depends on the oblateness of the star.   \cite{katz17} applies  the precession model  to explain the orbital waveform in the optical bands, and predicts a  precession period of 20-200 years for \src. \cite{tovmassian07, tovmassian12}  report an evidence of the WD's precession in the X-ray emission from two IPs. If this precession scenario is  true for \src,  the transition between a single-peak and a double-peak structures in the pulse profile folded with the beat frequency will be regularly repeated with the precession frequency. Since the pulse peak in the X-ray bands is aligned with the peak in the optical/UV bands, more long-term multi-wavelength observations
 will be required  to understand the origin of the long-term evolution of the emission from \src.

We thank to referee for his/her useful comments and suggestions. J.T., W.X.F and  W.H.H. are supported by the National Science Foundation of China (NSFC) under 11573010, 11661161010, U1631103 and U1838102. L.C.C.L. is supported by NRFK through the grant number of 2016R1A5A1013277.  K.L.L is supported by the Ministry of Science and Technology of the Republic of China (Taiwan) through grants 108-2112-M-007-025-MY3 and 109-2636-M-006-017, and he is a Yushan (Young) Scholar of the Ministry of Education of the Republic of China (Taiwan). AKHK is supported by the Ministry of Science and Technology of the Republic of China (Taiwan) through grants 108-2628-M-007-005-RSP and 109-2628-M-007-005-RSP.

\bibliography{adssample}

\begin{thebibliography}{}
\expandafter\ifx\csname natexlab\endcsname\relax\def\natexlab#1{#1}\fi

\bibitem[{{Bailer-Jones} {et~al.}(2018){Bailer-Jones}, {Rybizki}, {Fouesneau},
  {Mantelet}, \& {Andrae}}]{bailer18}
{Bailer-Jones}, C.~A.~L., {Rybizki}, J., {Fouesneau}, M., {Mantelet}, G., \&
  {Andrae}, R. 2018, \aj, 156, 58

\bibitem[{{Bednarek}(2018)}]{bednarek18}
{Bednarek}, W. 2018, \mnras, 476, L10

\bibitem[{{Buccheri} {et~al.}(1983){Buccheri}, {Bennett}, {Bignami}, {Bloemen},
  {Boriakoff}, {Caraveo}, {Hermsen}, {Kanbach}, {Manchester}, {Masnou},
  {Mayer-Hasselwander}, {{\"O}zel}, {Paul}, {Sacco}, {Scarsi}, \&
  {Strong}}]{buccheri83}
{Buccheri}, R., {Bennett}, K., {Bignami}, G.~F., {et~al.} 1983, \aap, 128, 245

\bibitem[{{Buckley} {et~al.}(2017){Buckley}, {Meintjes}, {Potter}, {Marsh}, \&
  {G{\"a}nsicke}}]{buckley17}
{Buckley}, D.~A.~H., {Meintjes}, P.~J., {Potter}, S.~B., {Marsh}, T.~R., \&
  {G{\"a}nsicke}, B.~T. 2017, Nature Astronomy, 1, 0029

\bibitem[{{Davis}(2001)}]{davis01}
{Davis}, J.~E. 2001, \apj, 562, 575

\bibitem[{{du Plessis} {et~al.}(2019){du Plessis}, {Wadiasingh}, {Venter}, \&
  {Harding}}]{duplessis19}
{du Plessis}, L., {Wadiasingh}, Z., {Venter}, C., \& {Harding}, A.~K. 2019,
  \apj, 887, 44

\bibitem[{{Gaibor} {et~al.}(2020){Gaibor}, {Garnavich}, {Littlefield},
  {Potter}, \& {Buckley}}]{gaibor20}
{Gaibor}, Y., {Garnavich}, P.~M., {Littlefield}, C., {Potter}, S.~B., \&
  {Buckley}, D.~A.~H. 2020, \mnras, 496, 4849

\bibitem[{{Garnavich} {et~al.}(2019){Garnavich}, {Littlefield}, {Kafka},
  {Kennedy}, {Callanan}, {Balsara}, \& {Lyutikov}}]{garnavich19}
{Garnavich}, P., {Littlefield}, C., {Kafka}, S., {et~al.} 2019, \apj, 872, 67

\bibitem[{{Geng} {et~al.}(2016){Geng}, {Zhang}, \& {Huang}}]{geng16}
{Geng}, J.-J., {Zhang}, B., \& {Huang}, Y.-F. 2016, \apjl, 831, L10

\bibitem[{{Hameury} \& {Lasota}(2017)}]{hameury17}
{Hameury}, J.~M., \& {Lasota}, J.~P. 2017, \aap, 602, A102

\bibitem[{{Hellier} {et~al.}(1997){Hellier}, {Mukai}, \&
  {Beardmore}}]{hellier97}
{Hellier}, C., {Mukai}, K., \& {Beardmore}, A.~P. 1997, \mnras, 292, 397

\bibitem[{{{\.I}nam} {et~al.}(2004){{\.I}nam}, {Baykal}, {Matthew Scott},
  {Finger}, \& {Swank}}]{inam04}
{{\.I}nam}, S.~{\c{C}}., {Baykal}, A., {Matthew Scott}, D., {Finger}, M., \&
  {Swank}, J. 2004, \mnras, 349, 173

\bibitem[{{Kaplan} {et~al.}(2019){Kaplan}, {Meintjes}, {Singh}, {van Heerden},
  {Ramamonjisoa}, \& {van der Westhuizen}}]{kaplan19}
{Kaplan}, Q., {Meintjes}, P.~J., {Singh}, K.~K., {et~al.} 2019, arXiv e-prints,
  arXiv:1908.00283

\bibitem[{{Katz}(2017)}]{katz17}
{Katz}, J.~I. 2017, \apj, 835, 150

\bibitem[{{Littlefield} {et~al.}(2017){Littlefield}, {Garnavich}, {Kennedy},
  {Callanan}, {Shappee}, \& {Holoien}}]{littlefield17}
{Littlefield}, C., {Garnavich}, P., {Kennedy}, M., {et~al.} 2017, \apjl, 845,
  L7

\bibitem[{{Lyutikov} {et~al.}(2020){Lyutikov}, {Barkov}, {Route}, {Balsara},
  {Garnavich}, \& {Littlefield}}]{lyutikov20}
{Lyutikov}, M., {Barkov}, M., {Route}, M., {et~al.} 2020, arXiv e-prints,
  arXiv:2004.11474

\bibitem[{{Marcote} {et~al.}(2017){Marcote}, {Marsh}, {Stanway}, {Paragi}, \&
  {Blanchard}}]{marcote17}
{Marcote}, B., {Marsh}, T.~R., {Stanway}, E.~R., {Paragi}, Z., \& {Blanchard},
  J.~M. 2017, \aap, 601, L7

\bibitem[{{Marsh} {et~al.}(2016){Marsh}, {G{\"a}nsicke}, {H{\"u}mmerich},
  {Hambsch}, {Bernhard}, {Lloyd}, {Breedt}, {Stanway}, {Steeghs}, {Parsons},
  {Toloza}, {Schreiber}, {Jonker}, {van Roestel}, {Kupfer}, {Pala}, {Dhillon},
  {Hardy}, {Littlefair}, {Aungwerojwit}, {Arjyotha}, {Koester}, {Bochinski},
  {Haswell}, {Frank}, \& {Wheatley}}]{marsh16}
{Marsh}, T.~R., {G{\"a}nsicke}, B.~T., {H{\"u}mmerich}, S., {et~al.} 2016,
  \nat, 537, 374

\bibitem[{{Mewe} {et~al.}(1985){Mewe}, {Gronenschild}, \& {van den
  Oord}}]{mewe85}
{Mewe}, R., {Gronenschild}, E.~H.~B.~M., \& {van den Oord}, G.~H.~J. 1985,
  \aaps, 62, 197

\bibitem[{{Mewe} {et~al.}(1986){Mewe}, {Lemen}, \& {van den Oord}}]{mewe86}
{Mewe}, R., {Lemen}, J.~R., \& {van den Oord}, G.~H.~J. 1986, \aaps, 65, 511

\bibitem[{{Norton} \& {Mukai}(2007)}]{norton07}
{Norton}, A.~J., \& {Mukai}, K. 2007, \aap, 472, 225

\bibitem[{{Peterson} {et~al.}(2019){Peterson}, {Littlefield}, \&
  {Garnavich}}]{peterson19}
{Peterson}, E., {Littlefield}, C., \& {Garnavich}, P. 2019, \aj, 158, 131

\bibitem[{{Potter} \& {Buckley}(2018{\natexlab{a}})}]{potter18a}
{Potter}, S.~B., \& {Buckley}, D. A.~H. 2018{\natexlab{a}}, \mnras, 478, L78

\bibitem[{{Potter} \& {Buckley}(2018{\natexlab{b}})}]{potter18b}
---. 2018{\natexlab{b}}, \mnras, 481, 2384

\bibitem[{{Rea} {et~al.}(2007){Rea}, {Nichelli}, {Israel}, {Perna},
  {Oosterbroek}, {Parmar}, {Turolla}, {Campana}, {Stella}, {Zane}, \&
  {Angelini}}]{rea07}
{Rea}, N., {Nichelli}, E., {Israel}, G.~L., {et~al.} 2007, \mnras, 381, 293

\bibitem[{{Schwarzenberg-Czerny}(1992)}]{schwarzenberg92}
{Schwarzenberg-Czerny}, A. 1992, \aap, 260, 268

\bibitem[{{Singh} {et~al.}(2020){Singh}, {Meintjes}, {Kaplan}, {Ramamonjisoa},
  \& {Sahayanathan}}]{singh20}
{Singh}, K.~K., {Meintjes}, P.~J., {Kaplan}, Q., {Ramamonjisoa}, F.~A., \&
  {Sahayanathan}, S. 2020, Astroparticle Physics, 123, 102488

\bibitem[{{Stanway} {et~al.}(2018){Stanway}, {Marsh}, {Chote}, {G{\"a}nsicke},
  {Steeghs}, \& {Wheatley}}]{stanway18}
{Stanway}, E.~R., {Marsh}, T.~R., {Chote}, P., {et~al.} 2018, \aap, 611, A66

\bibitem[{{Stiller} {et~al.}(2018){Stiller}, {Littlefield}, {Garnavich},
  {Wood}, {Hambsch}, \& {Myers}}]{stiller18}
{Stiller}, R.~A., {Littlefield}, C., {Garnavich}, P., {et~al.} 2018, \aj, 156,
  150

\bibitem[{{Takata} \& {Cheng}(2019)}]{takata19}
{Takata}, J., \& {Cheng}, K.~S. 2019, \apj, 875, 119

\bibitem[{{Takata} {et~al.}(2018){Takata}, {Hu}, {Lin}, {Tam}, {Pal}, {Hui},
  {Kong}, \& {Cheng}}]{takata18}
{Takata}, J., {Hu}, C.~P., {Lin}, L.~C.~C., {et~al.} 2018, \apj, 853, 106

\bibitem[{{Takata} {et~al.}(2017){Takata}, {Yang}, \& {Cheng}}]{takata17}
{Takata}, J., {Yang}, H., \& {Cheng}, K.~S. 2017, \apj, 851, 143

\bibitem[{{Tovmassian} {et~al.}(2012){Tovmassian}, {Zharikov}, \&
  {Neustroev}}]{tovmassian12}
{Tovmassian}, G., {Zharikov}, S., \& {Neustroev}, V. 2012, \memsai, 83, 713

\bibitem[{{Tovmassian} {et~al.}(2007){Tovmassian}, {Zharikov}, \&
  {Neustroev}}]{tovmassian07}
{Tovmassian}, G.~H., {Zharikov}, S.~V., \& {Neustroev}, V.~V. 2007, \apj, 655,
  466

\end{thebibliography}
\end{document}